\documentclass[9pt,journal]{IEEEtran}
\pdfoutput=1 % if your are submitting a pdflatex (i.e. if you have
\usepackage{graphicx,bm,epsf,float,amssymb}
\usepackage[font={scriptsize}]{subcaption} % get subfigures
\usepackage{xcolor}
\usepackage{lineno}
\usepackage{multirow}
\usepackage{xspace}
\usepackage{url}
\usepackage[numbers,sort&compress]{natbib}
\usepackage{etoolbox}
\apptocmd{\sloppy}{\hbadness 10000\relax}{}{}

\usepackage{siunitx}
\usepackage{array}
\usepackage{amsmath}
\usepackage{xspace}
\newcommand{\fig}[1]{Fig.~\ref{#1}}
\newcommand{\Fig}[1]{Figure~\ref{#1}}

\newcommand{\En}{\ensuremath{E_{n}}\xspace}
\newcommand{\Enp}{\ensuremath{E_{n'}}\xspace}
\newcommand{\Natt}{\ensuremath{ ^{22}\text{Na}}\xspace}

\begin{document}
% \begin{frontmatter}
\title{Design and Characterization of an Optically-Segmented Single Volume Scatter Camera Module}

\author{
  Kevin Keefe, % Software, Formal analysis, Investigation, Data curation, writing-original draft
  Hassam Alhajaji, % Resources
  Erik Brubaker, % Conceptualization, Writing - Review & Editing, Supervision, Funding acquisition
  Andrew Druetzler, % Methodology, Investigation, Resources, Writing-Review and Editing, Supervision, Project Administration.
  Aline Galindo-Tellez, % Writing-Review and Editing
  John Learned, % Writing-Review and Editing
  Paul Maggi, % Writing-Review and Editing, Validation, Investigation
  Juan J.Manfredi, % Writing - Review and Editing, Software,  Data Curation,  Visualization
  Kurtis Nishimura, \IEEEmembership{Member, IEEE}, % Conceptualization, Resources, Writing, review & editing, Supervision, Project administration, Funding acquisition
  Bejamin Pinto Souza, % Resources
  John Steele, % Resources
  Melinda Sweany, % Conceptualization, Methodology, Validation, Investigation, Writing-Original Draft, Writing-Review and Editing, Supervision, Project administration
  Eric Takahashi % Resources
  \thanks{
    Kevin Keefe (email: kevinpk@hawaii.edu), Hassam Alhajaji, Andrew Druetzler, John Learned, Kurtis
    Nishimura, Bejamin Pinto Souza, and Eric Takahashi are with University of
    Hawaii, Manoa HI 96826 USA.

    Aline Galindo-Tellez was with University of Hawaii, Manoa HI 96826 USA and
    is now with Institut de recherche sur les lois Fondamentales de l'Univers,  CEA,
    Université Paris-Sacley, 91191.

    Erik Brubaker, Melinda Sweany, and John Steele are with Sandia National
    Laboratories, CA 94550.

    Paul Maggi was with with Sandia National Laboratories, CA 94550 USA
    and is now with Lawrence Livermore National Laboratories 94550 USA.

    Juan Manfredi was with the University of Berkeley, CA 94720 USA and is now with
    Air Force Institute of Technology at Wright-Patterson Air Force Base
    in Ohio 45433 USA.
  }
}
\maketitle

\begin{abstract}
The Optically Segmented Single Volume Scatter Camera (OS-SVSC) aims to image
neutron sources for nuclear non-proliferation applications using the kinematic
reconstruction of elastic double-scatter events. We report on the design,
construction, and calibration of one module of a new prototype. The
module includes 16 EJ-204 organic plastic scintillating bars individually
wrapped in Teflon tape, each measuring 0.5~cm$\times$0.5~cm$\times$20~cm.
The scintillator array is coupled to two custom Silicon
Photomultiplier (SiPM) boards consisting of a 2$\times$8 array of SensL
J-Series-60035 Silicon Photomultipliers, which are read out by a custom 16
channel DRS-4 based digitizer board. The electrical crosstalk between SiPMs
within the electronics chain is measured as 0.76\% $\pm$ 0.11\%
among all 16 channels. We
report the detector response of one module including interaction position, time,
and energy, using two different optical coupling materials: EJ-560 silicone
rubber optical coupling pads and EJ-550 optical coupling grease. We present
results in terms of the overall mean and standard deviation of the z-position
reconstruction and interaction time resolutions for all 16 bars in the module.
We observed the z-position resolution for gamma interactions in the
0.3 MeVee to 0.4 MeVee range to be 2.24~cm$\pm$1.10~cm and
1.45~cm$\pm$0.19~cm for silicone optical coupling pad and optical
grease, respectively. The observed interaction time resolution is
265~ps$\pm$29~ps and 235~ps$\pm$10~ps for silicone
optical coupling pad and optical grease, respectively.
\end{abstract}


% \end{frontmatter}

%------------------------------------------------------------------------------------------------------------------------
\section{Introduction}

The Single Volume Scatter Camera (SVSC) project aims to develop a portable
kinematic neutron imaging system~\cite{Manfredi}. In comparison to the conventional
implementation~\cite{Sailor1991,Mascarenhas2009,Madden2013,Goldsmith2016,DPI}, 
a compact scintillator
geometry is used to reconstruct multiple neutron interactions, rather than
distributed volumes. It is estimated that a single-volume neutron scatter camera
with an interaction resolution of $O(\SI{1}{cm},\SI{1}{ns})$ can result in an 
order of magnitude improvement in detection efficiency compared to
traditional designs~\cite{svsc}, in addition
to improvements in size, weight, and power of the instrument.
Reduced system size enhances the practicality of the device in the field and
also enables additional performance boosts via closer physical distance to
objects of interest.

The SVSC relies on accurately reconstructing the time, position, and energy of
two neutron-proton interactions within the compact volume of scintillator. The
reconstruction of the incoming neutron direction is dependent on the incoming
neutron energy, \En, and the energy after the first interaction \Enp:

\begin{equation}
\mathrm{\cos}(\theta) = \sqrt{\frac{\Enp}{\En}}.
\end{equation}

\Enp is determined by the neutron time-of-flight equation, requiring
knowledge of the position and time of the first two neutron-proton interactions:

\begin{equation}
\Enp = \frac{1}{2} m_n \left(\frac{\Delta d}{\Delta t}\right)^{2}.
\end{equation}

Finally, \En is determined by the sum of \Enp and the recoil energy of the
proton from the first interaction, which is found using the proton light yield
relation of the scintillator material~\cite{lightYield}.

Monolithic compact designs propose to reconstruct multiple interaction
positions within a single block of scintillator~\cite{svsc,mTC}. In contrast,
optically segmented (OS)
approaches~\cite{Weinfurther,galindo-tellez,Steinberger2020,Wonders2021,Pang2019}
employ an array of rectangular scintillator bars that are read out on both
sides: for each interaction, the time and position along the bar are
reconstructed, while the other two dimensions are determined by which bar the
scintillation event occurred in. Current methods to reconstruct the position
along the bar include the log of the ratio of amplitudes at the two readout ends
and the difference in pulse time arrival at the two ends. The interaction time
is measured with the average pulse time arrival at the two ends. The
performance of these position and time reconstruction methods depends on  the
properties of the scintillator and any reflective
wrapping~\cite{Sweany,Steinberger2019}.

Based on our study of different scintillator materials and reflective wrappings,
we previously constructed a prototype OS-SVSC system with 
$\SI{0.5}{cm}\times\SI{0.5}{cm}\times\SI{20}{cm}$ EJ-204 
scintillator bars wrapped in Teflon tape, read out
by two commercially available ArrayJ-60035-64P-PCB Silicon Photomultiplier
(SiPM) arrays from SensL. The full description of that system and its
performance are reported in~\cite{galindo-tellez}.

The prior system was affected by electrical crosstalk at the
\SI{10}{\percent} level within the electronics chain and the SiPM array itself,
which is expected to  negatively affect interaction reconstructions, as
crosstalk from one neutron  interaction could bias the time and energy
measurements of the other interaction. Further tests demonstrated that crosstalk
in the commercial SiPM array occurs primarily in quadrants, suggesting that one
large contributor to this crosstalk is the SAMTEC connector
(QTE-040-03)~\cite{galindo-tellez}.

The system was also likely affected by optical crosstalk between bars and
suboptimal light collection due to the \SI{0.5}{mm} thick EJ-560 silicone
rubber optical coupling pads. A \SI{0.5}{mm} stand-off between the
scintillator and the SiPMs, which are separated by \SI{0.2}{mm} within the
ArrayJ-60035-64P-PCB, is likely to cause some optical crosstalk, even with the
individually cut $\SI{5}{mm}\times\SI{5}{mm}$ silicone optical pads used in
the prototype. In addition, light escaping the edges of the pad can result in
light loss, leading to degradation of resolution. Misalignment of the bars due
to non-uniform Teflon wrapping is another potential source of optical crosstalk
and light loss.

Detailed characterization of each bar in the prior $8\times8$ 
array was a challenge. Prior
single-bar characterization of the interaction time, energy, and position resolution
were achieved with back-to-back \SI{0.511}{MeV} annihilation gamma rays from a
\Natt source: using a tag scintillator with a known position along the bar,
the position-dependent response was measured~\cite{Sweany}. However, for bars
near the center of a full array, this was difficult to achieve due to scattering in
the outside bars. Thus, the inner bars of the prototype were characterized by
using the outer bar position calibration to reconstruct individual cosmic muon
tracks within the array, a time-consuming process.

We designed a second prototype based on the lessons learned in deploying and
characterizing this prior prototype. Many of the engineering and detector
characterization difficulties have been addressed by employing a modular design
in which the detector can be split into multiple $2\times8$ scintillator arrays.
We designed a custom SiPM array, which allowed for extra care
in avoiding optical and electronic crosstalk: the pixels are separated by
\SI{2}{mm}, rather than \SI{0.2}{mm}, so that optical crosstalk is more easily avoided, and the circuit
was designed specifically to minimize electronic crosstalk between channels. The
board is designed to interface directly with the Sandia Laboratories Compact
Electronics for Modular Acquisition (SCEMA) electronics board~\cite{steele}. Finally,
the modular design assures that each bar is accessible for calibrations using a 
tagged \Natt source.

%******************************************************************************

\section{Module design}

\begin{figure}
\centering
  \begin{subfigure}{\columnwidth}
    \centering
    \includegraphics[width=\columnwidth]{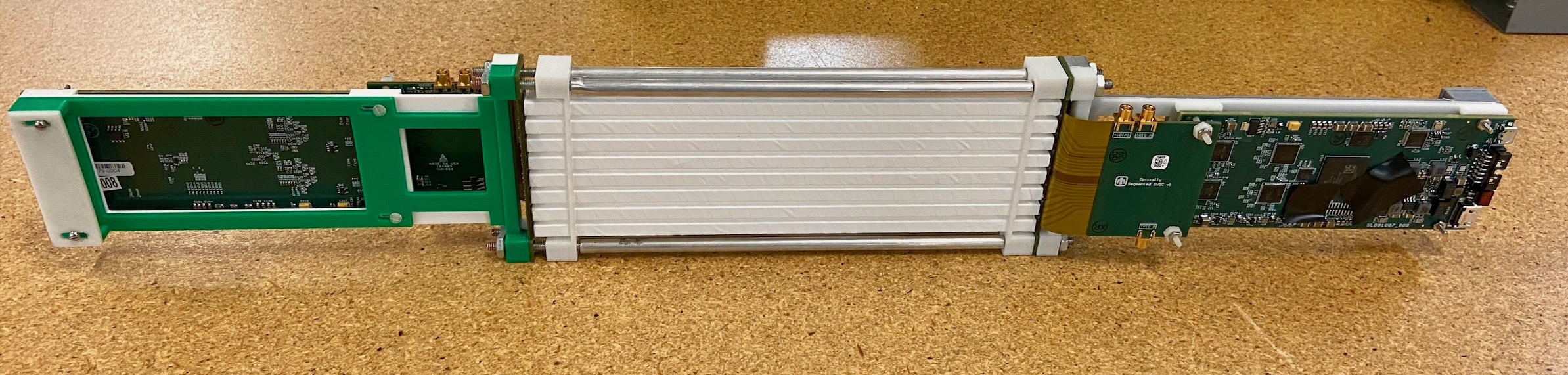}
    \caption{}
  \end{subfigure}
  \begin{subfigure}{\columnwidth}
    \centering
    \includegraphics[trim=0 216 0 400, clip, width=\columnwidth]{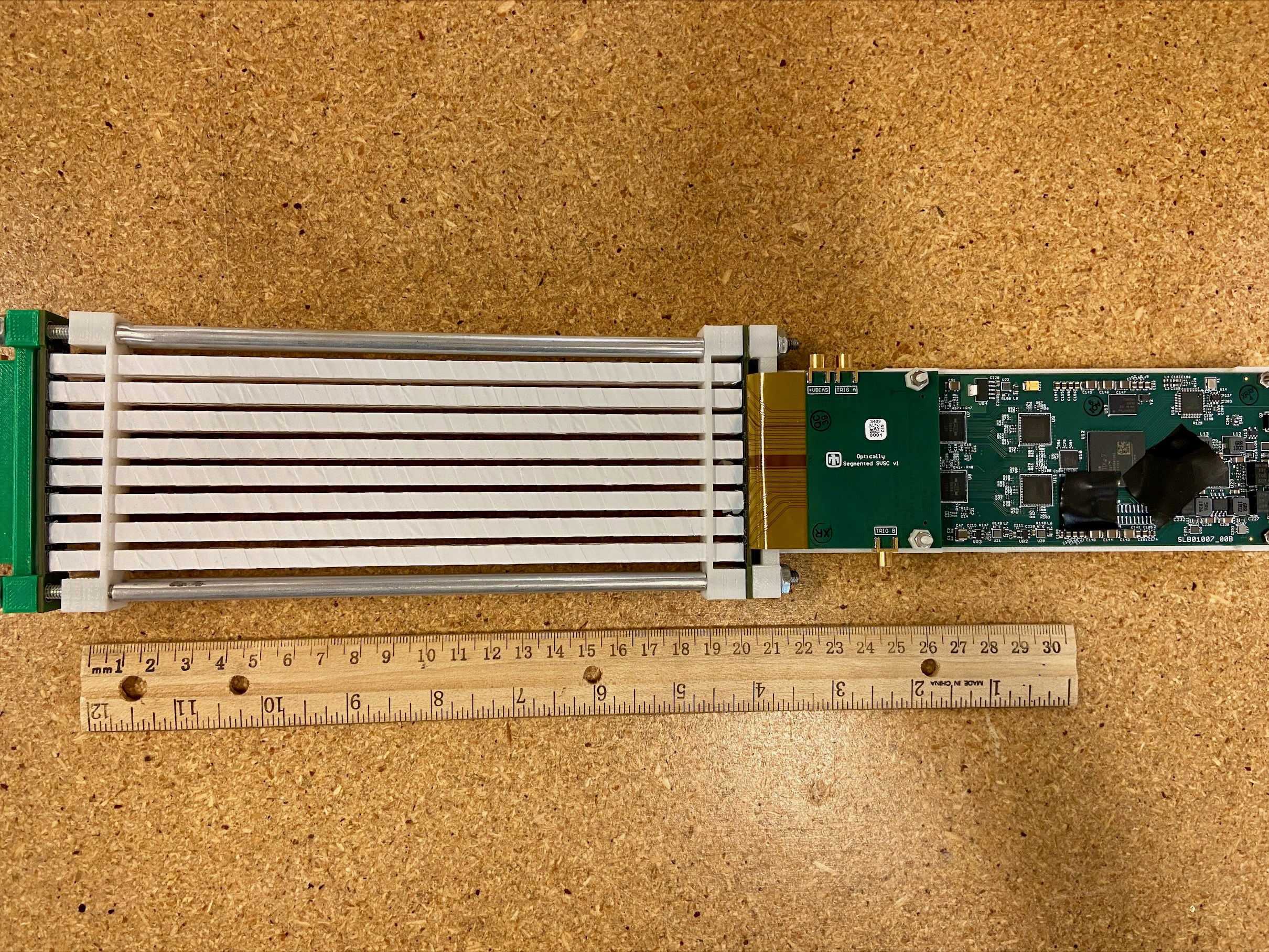}
    \caption{}
  \end{subfigure}
\caption{
(a) Perspective view of a fully populated optically segmented module (OSMO), 
with 16 Teflon-wrapped EJ-204 bars.
(b) Zoomed in photo of the OSMO for scale, lying flat on the table.
}
\label{fig:osmo}
\end{figure}

As stated above, the second OS-SVSC prototype uses a modular design to minimize 
electrical and optical crosstalk, and to enable faster and more accurate
detector characterizations. A single Optically Segmented MOdule (OSMO) includes 
a total of sixteen $\SI{0.5}{cm}\times\SI{0.5}{cm}\times\SI{20}{cm}$ EJ-204 
scintillator bars~\cite{ej204} configured in a $2\times8$ array, as well as 
associated readout electronics. One such OSMO is shown in~\fig{fig:osmo}.
In this section we describe the specifics of the design, including
the mechanical features and front-end electronics.

%*******************************************************************************
\subsection{Mechanical design}

Each scintillator bar is wrapped with at minimum three layers of white Teflon
tape.  The ends of each bar are coupled to SiPMs using either EJ-560 silicone
rubber optical coupling pads~\cite{ej560}  or EJ-550 optical
grease~\cite{ej550}; we report results for both coupling materials.

The scintillator bars are fed through 3-D printed plastic grids for support and
alignment. In order to provide the pressure between the bars and the SiPMs for
optimal optical coupling, we use four threaded aluminum rods.
Screws pass through the plastic guides and a plastic 3-D printed support
bracket. This support bracket provides the base for each of the readout boards to
be rigidly connected to both the bar support guide and the threaded rods.
Finally, two sets of plastic 3-D printed grids ensure that the
optical pads and bars are aligned with the SiPMs and provide optical isolation
between adjacent channels.

%*******************************************************************************
\subsection{Front-end electronics design and Data Acquisition}

A new custom readout system was developed for the OSMO, based closely on a
previous design~\cite{steele}. The electronics components consist of two
interconnected PCBs: the SiPM module with a $2\times8$ SiPM array and the 
SCEMA waveform capture board.

The SCEMA-B PCB used for all results in this paper is based on the original
SCEMA-A design~\cite{steele}, with modifications made to improve FPGA IO mappings,
address an existing ADC problem on one channel, and to split functionality to
provide a modular front-end.  Whereas the SCEMA-A had a front-end that was
designed to directly connect to a Planacon MCP-PMT header, the SCEMA-B includes
a Samtec connector that allows for implementation of various front-ends,
including a similar Planacon interface board, but also allowing for other custom
readouts, such as the SiPM module described herein.

\begin{figure}
\centering
  \begin{subfigure}{0.49\columnwidth}
    \centering
    \includegraphics[width=\columnwidth]{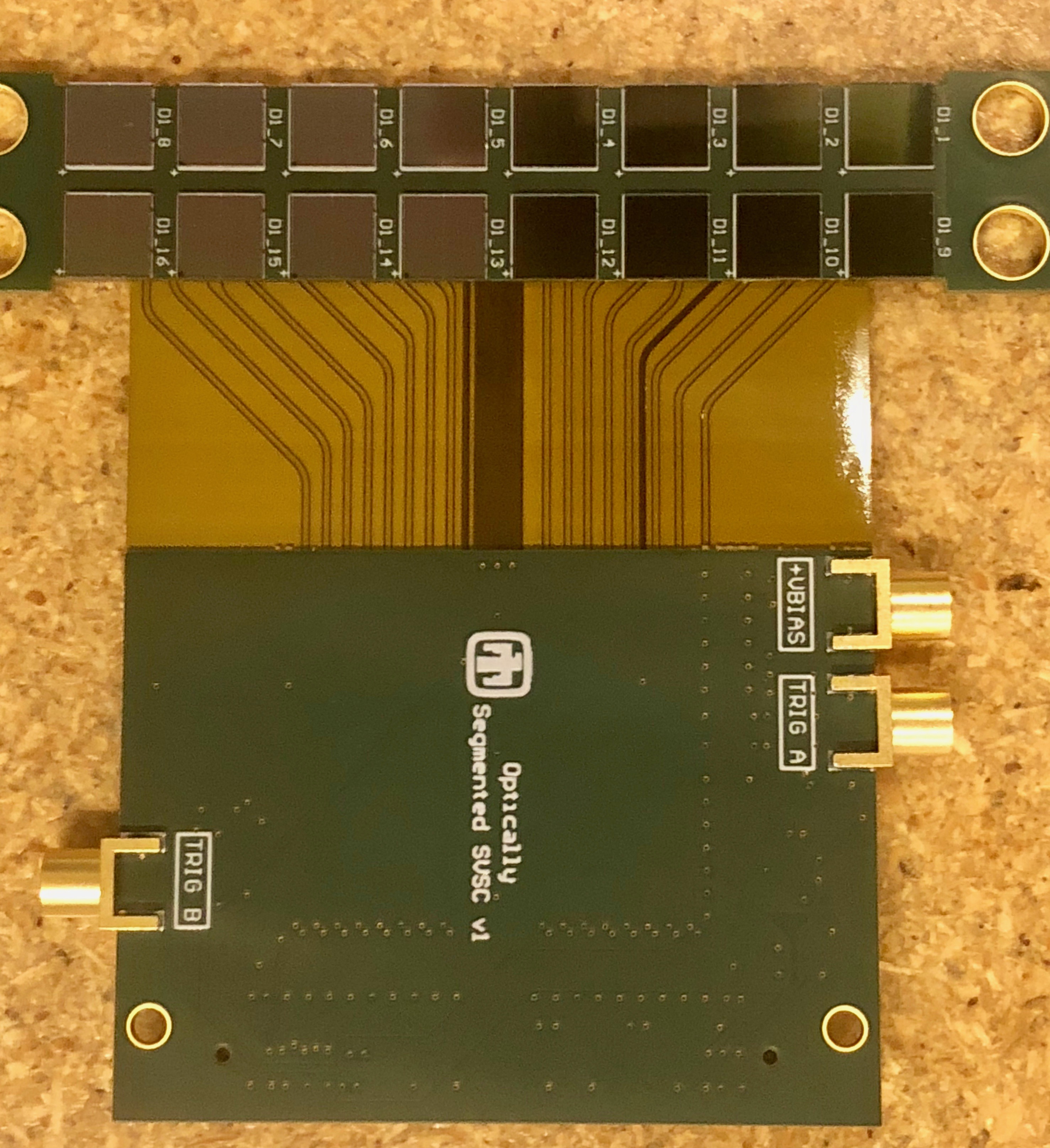}
    \caption{}
  \end{subfigure}
  \begin{subfigure}{0.49\columnwidth}
    \centering
    \includegraphics[width=\columnwidth]{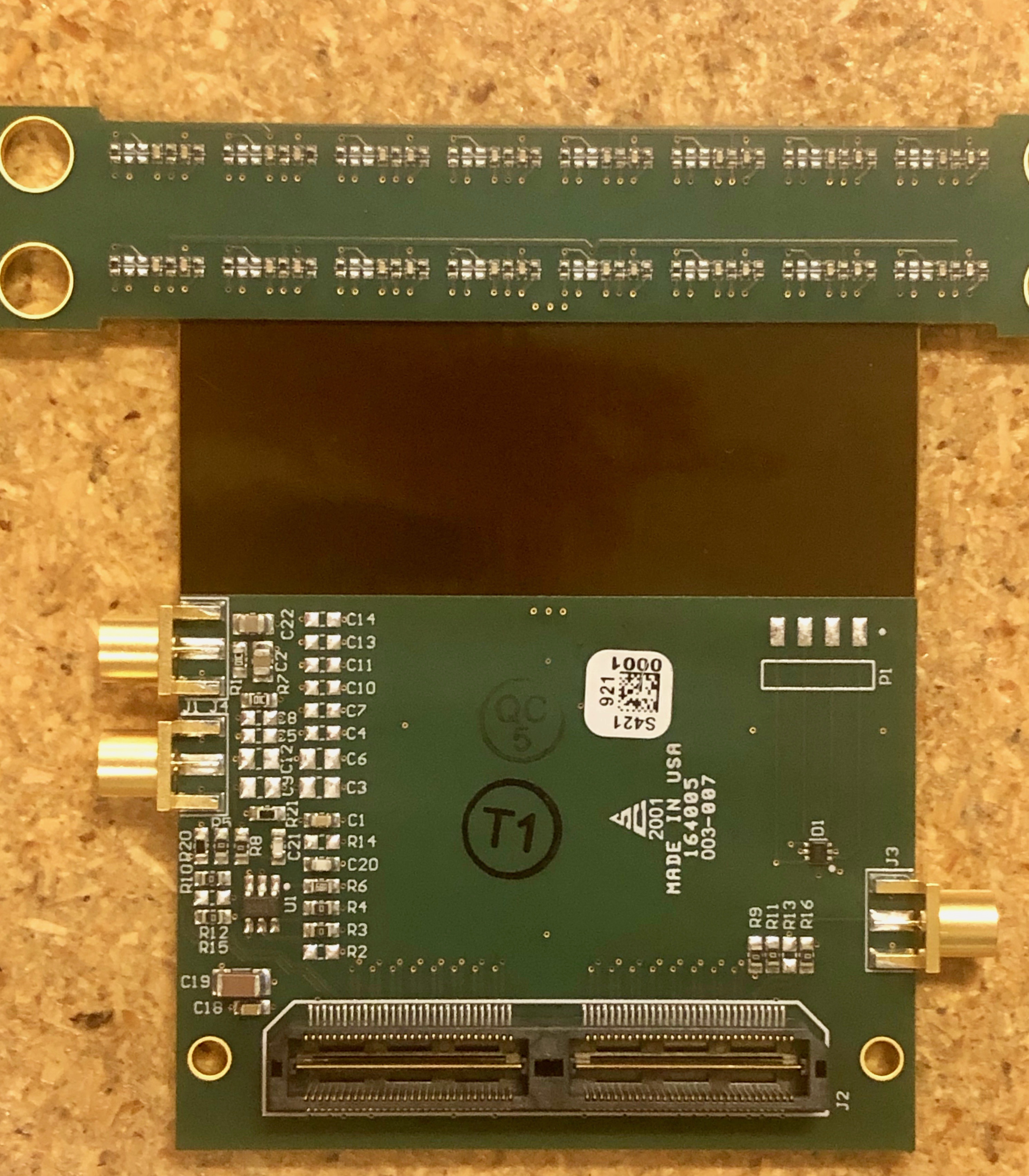}
    \caption{}
\end{subfigure}
\caption{The top (a) and bottom (b) sides of the $2\times8$ OSMO SiPM module 
board.}
\label{fig:interposer_photo}
\end{figure}

The SiPM module, shown in \fig{fig:interposer_photo}, consists of 16 SensL
J-Series SiPMs (MicroFJ-60035-TSV), which have a 
$\SI{6.13}{mm}\times\SI{6.13}{mm}$ cross-section and are 
separated by \SI{2}{mm}. A flex section
allows two separate rigid PCBs to couple in an orthogonal orientation without an
extra connector. The SiPM module is responsible for electrically
isolating each of the 16 SiPM anodes in the array.

The SiPM module connects to the SCEMA-B via a SamTec connector (QTH-060-01-L-D-A).
On each SCEMA-B are two DRS4 digitizers~\cite{Bitossi}. Each DRS4 has nine channels: eight of those channels digitize
half of the SiPM array and the remaining channel is responsible for digitizing
one of the trigger inputs. Thus, the SiPM module routes a total of 18 channels to
the SCEMA-B\@: 16 SiPM channels and 2 trigger channels, which we call Trigger-A
and Trigger-B.

The SiPM module also provides MCX connectors for the SiPM bias voltage, which is
supplied for these studies by a B\&K Precision 9129B DC power supply~\cite{bk9219}, and for the
two trigger channels. Trigger-A can be configured as either an external trigger 
(received via the MCX connector) or as a local sum trigger, depending on 
whether or not a summing OpAmp is populated. This sum trigger provides 
self-triggering capability for one OSMO for acquiring neutron scattering data.
The summing circuit reads a common-cathode signal of all 16 SiPMs and routes its
output both to a DRS4 channel for digitization and to the Trigger-A MCX
connector for output. Trigger-B is configured only as an external trigger; the
signal from its MCX connector is simply routed to the SCEMA-B for digitization.
Trigger-B is used to input the trace from the tag scintillator for the \Natt
calibrations described below, and can also be used to accept a trigger from the
other side of the array or from the entire prototype for neutron scatter data.

\begin{figure}
\centering
\includegraphics[width=0.98\columnwidth]{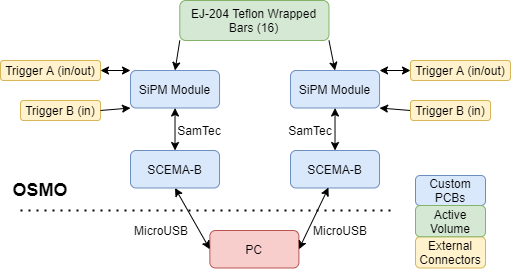}
\caption{Diagram of the OSMO DAQ configuration. }
\label{fig:osmo_daq}
\end{figure}

The OSMO's modular design has at least two different acquisition configurations
for the SCEMA-B's. Each SCEMA-B is able to run independently for module calibrations
as well as in a connected state in a multiple OSMO system. This design allows
for OSMOs to be assembled and calibrated individually, and then for multiple
OSMOs to be connected together to increase the active scintillator volume of a
full system. All of the data presented here was collected with each SCEMA-B
running independently and with Trigger-A configured as an input trigger. The
digitized data are sent from each SCEMA-B to a DAQ computer through a USB-2.0 
connection, and are then written to disk.
\Fig{fig:osmo_daq} summarizes the DAQ configuration.

%****************************************************************************************
\section{Module response characterization}

In this section we describe the full response of a single OSMO,
including electrical crosstalk on the SiPM module, SiPM module timing, and 
scintillator interaction response calibrations for the energy, position, and 
time. Results are presented for two scintillator/SiPM coupling materials: 
EJ-550 optical grease and \SI{0.5}{mm} thick EJ-560 silicone rubber pads.

%****************************************************************************************
\subsection{Waveform analysis}
\label{ss:waveformAnalysis}

Digitized waveforms from the SCEMA-B boards are analyzed with custom C++-based
software utilizing ROOT libraries~\cite{ROOT}. The values of interest are the
rising edge time of the trigger channels, the rising edge time of the SiPM
traces, and the maximum pulse height for each SiPM waveform. The DRS4s are sampled at
$\approx$\SI{4.9}{GHz}, and all 1024 bins within the switched capacitor array
(SCA) are digitized for each triggered event. Electrical calibrations for
timing and voltage are calculated using a similar procedure to that reported
previously~\cite{steele}.

It was also observed that there is some time dependent drift in the voltage
calibration for each of the bins within the DRS4 SCA\@. In order to account for
the voltage drift before each run, we subtract from each voltage bin the mean
average of the voltage of the same bin within the SCA\@. Before additional
processing but after DRS4 electrical calibrations, baseline offsets for the
entire waveform are calculated as the mean of the samples 20 through 60 of each
waveform.  After this procedure we obtain a voltage baseline deviation of
\SI{\approx1.0}{\micro V}, and voltage noise between \SIrange{0.3}{0.5}{mV}.

\begin{figure}
\centering
\includegraphics[width=0.98\columnwidth]{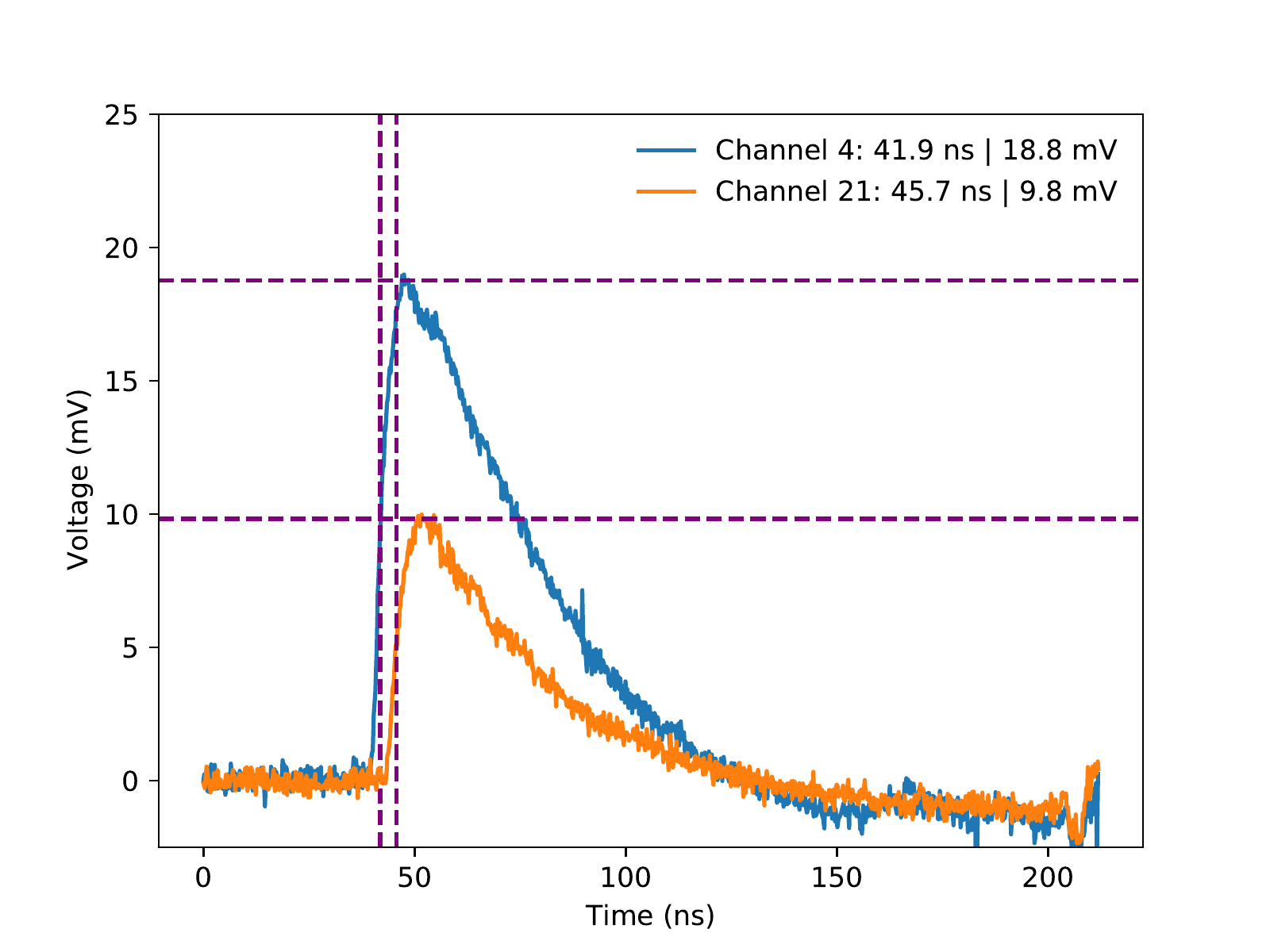}
\caption{An example bar interaction from a \Natt source with digitized 
         waveforms for the two SiPMs coupled to the ends of one bar. 
         The vertical lines represent
         where the analysis identifies the half-point rising edge of the waveform. The
         horizontal lines indicate the value the analysis identifies as the pulse
         height of the waveform.}
\label{fig:exampleBarEvent}
\end{figure}

Pulse heights of each SiPM trace are measured based on the following algorithm.
We first determine the six consecutive samples that give the largest integrated
value. We discard the highest and lowest of these six values and report the
average of the four remaining values as the pulse height. By dropping the
highest and lowest bin voltage values, we avoid noise that appears in some DRS4
bins, which (when present) was always separated by 32 bins and could not be
resolved with voltage calibrations alone.

Rise time measurements are determined from the interpolated value between the
samples corresponding to \SI{50}{\percent} of the measured pulse height
previously discussed. \Fig{fig:exampleBarEvent} shows the traces from both SiPMs
for an example bar interaction. The measured pulse heights and rise times for
both traces are indicated by vertical and horizontal dashed lines.

%****************************************************************************************
\subsection{Electrical crosstalk characterization}

\begin{figure}
\centering
  \begin{subfigure}{\columnwidth}
    \centering
    \includegraphics[width=\columnwidth]{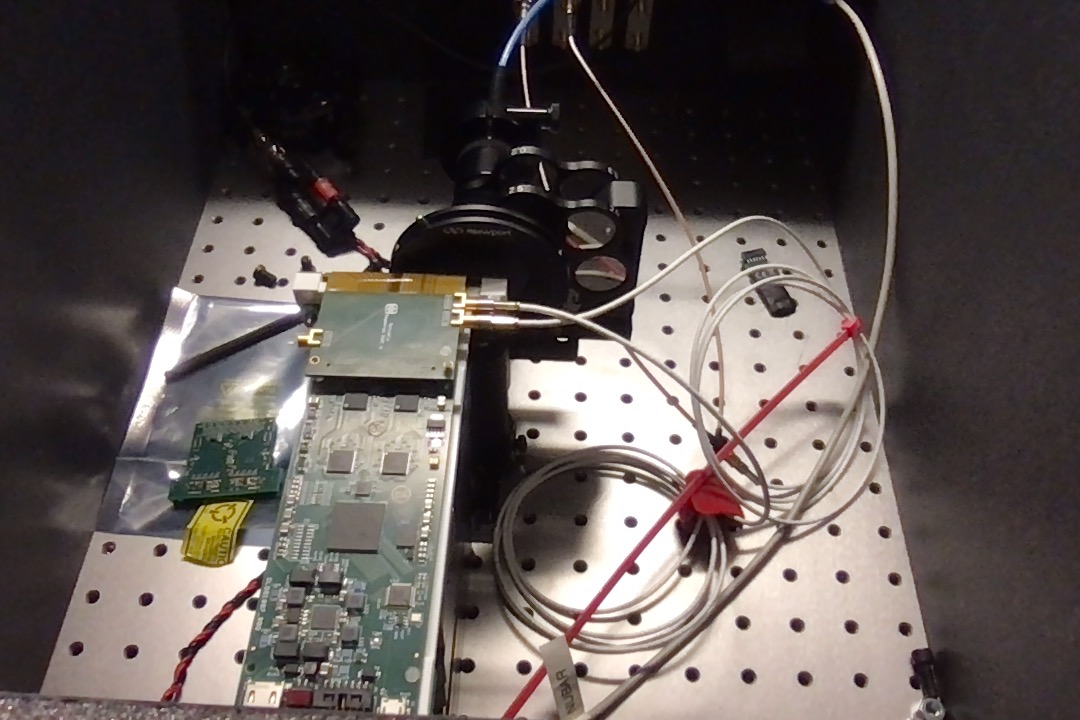}
    \caption{}
  \end{subfigure}
  \begin{subfigure}{\columnwidth}
    \centering
    \includegraphics[width=\columnwidth]{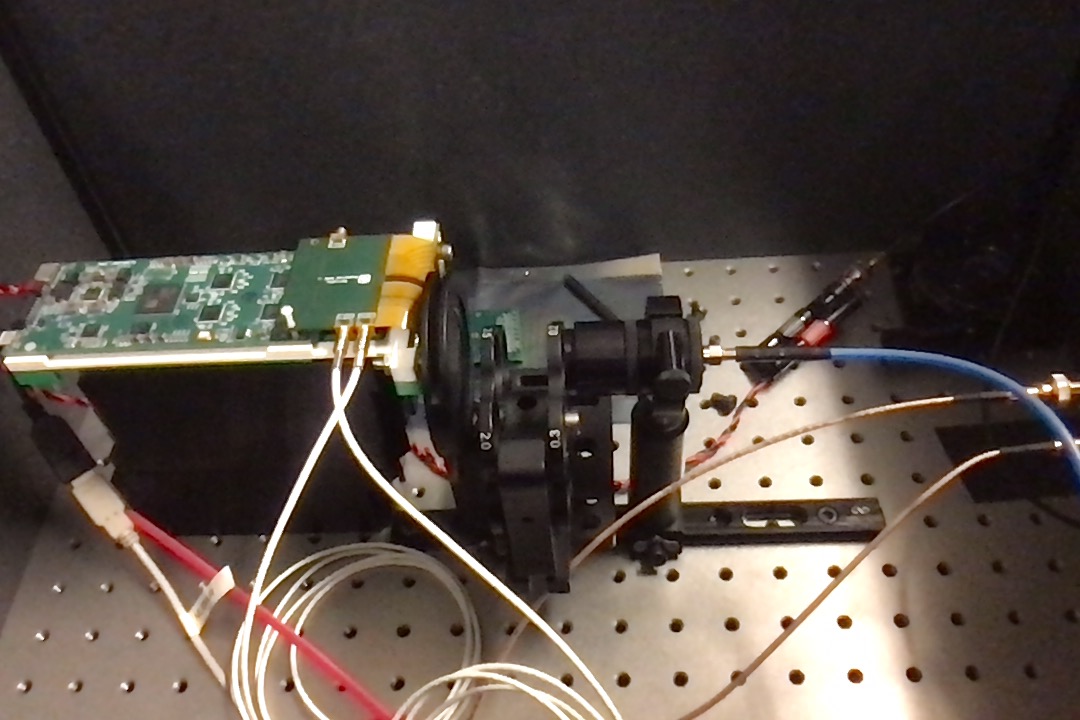}
    \caption{}
  \end{subfigure}
\caption{
(a) Image of the laser setup from behind. The whole SCEMA is aligned to the 
stationary optical shield to expose a single pixel to the laser.
(b) Image of the laser test-setup from the side. The blue cable is an optical 
fiber which directs the output laser packet. An optical shield and filter are 
used to isolate a single pixel as the target pixel for the run. The 
variable optical attenuator is used to tune the SiPM response to the laser.
}
\label{fig:laser_side}
\end{figure}

One motivation for a custom SiPM array was to reduce electrical crosstalk
between SiPM channels compared to the commercial J-series array~\cite{SensL2}.
To characterize the electrical crosstalk of the system, we used a Photek LPG-405
laser~\cite{lpg405} as a known input source on each of the 16 individual SiPM
channels, as shown in~\fig{fig:laser_side}. The laser was triggered externally
using a digital delay and pulse generator from Stanford Research Systems
(DG535), which was set to a nominal trigger rate of \SI{50}{Hz}, approximately
the maximum trigger rate for a single SCEMA in this setup. After the DG535 sends
a trigger to the laser, the laser then sends a trigger signal to Trigger-B, the
independent board trigger, followed $\approx$~\SI{47}{ns} later by an incident
photon packet. The resulting waveforms for all channels are then recorded. All
electrical crosstalk results are presented with the summing circuit populated.

We measure the amount of electrical crosstalk between a target channel and
remaining channels using both the coincident measured pulse height and the total
pulse integral of the target channel and the other 15 channels. The pulse height
measured for the remaining channels is the measured pulse height using the
algorithm defined in Section~\ref{ss:waveformAnalysis} where the measured pulse height must happen after
\SI{2}{ns} before the rising-edge of the target channel's response. This ensures
that the appropriate region of interest is measured. Similarly, the waveform
integral is calculated for all waveforms beginning \SI{2}{ns} before the
measured rising edge of the target channel.

%% example crosstalk calculation
\begin{figure}
\centering
\includegraphics[width=\columnwidth]{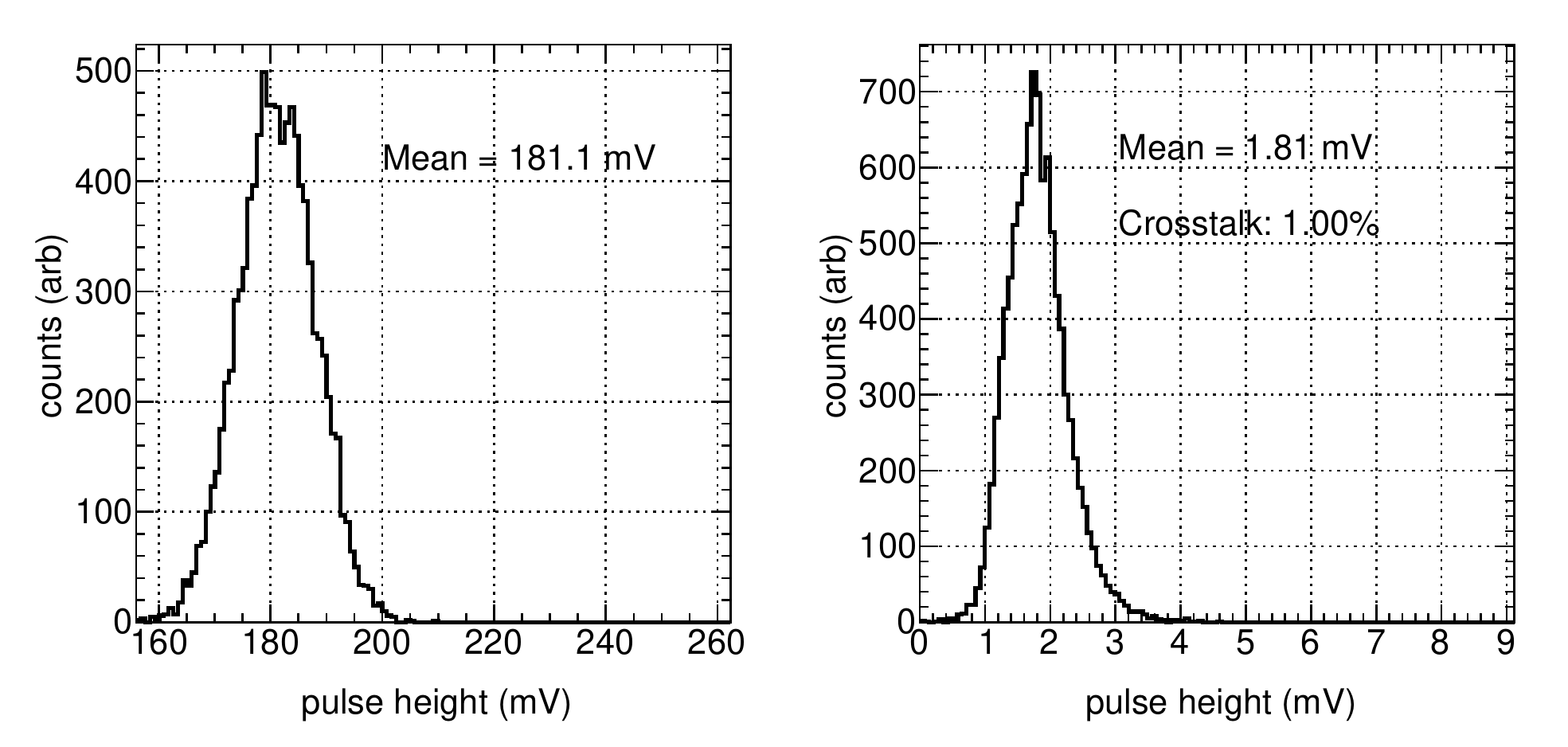}
\caption{Example pulse height measurement comparing the distribution of the
target channel (left) and an adjacent channel (right).}
\label{fig:crosstalkExample}
\end{figure}

%% OSMO 2 comparison figures
\begin{figure}
\centering
  \begin{subfigure}{\columnwidth}
    \centering
    \includegraphics[width=\columnwidth]{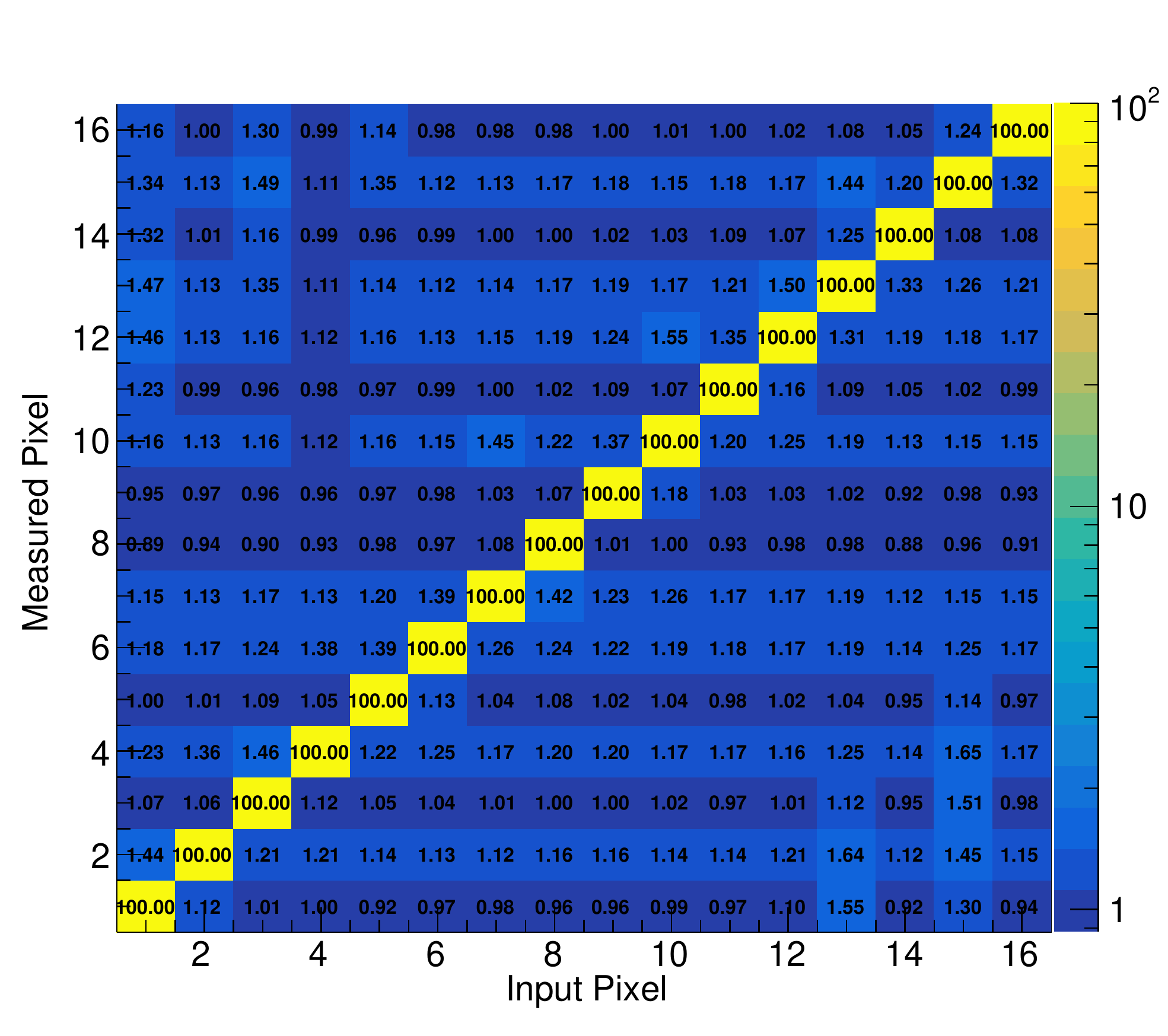}
    \caption{}
  \end{subfigure}
  \begin{subfigure}{\columnwidth}
    \centering
    \includegraphics[width=\columnwidth]{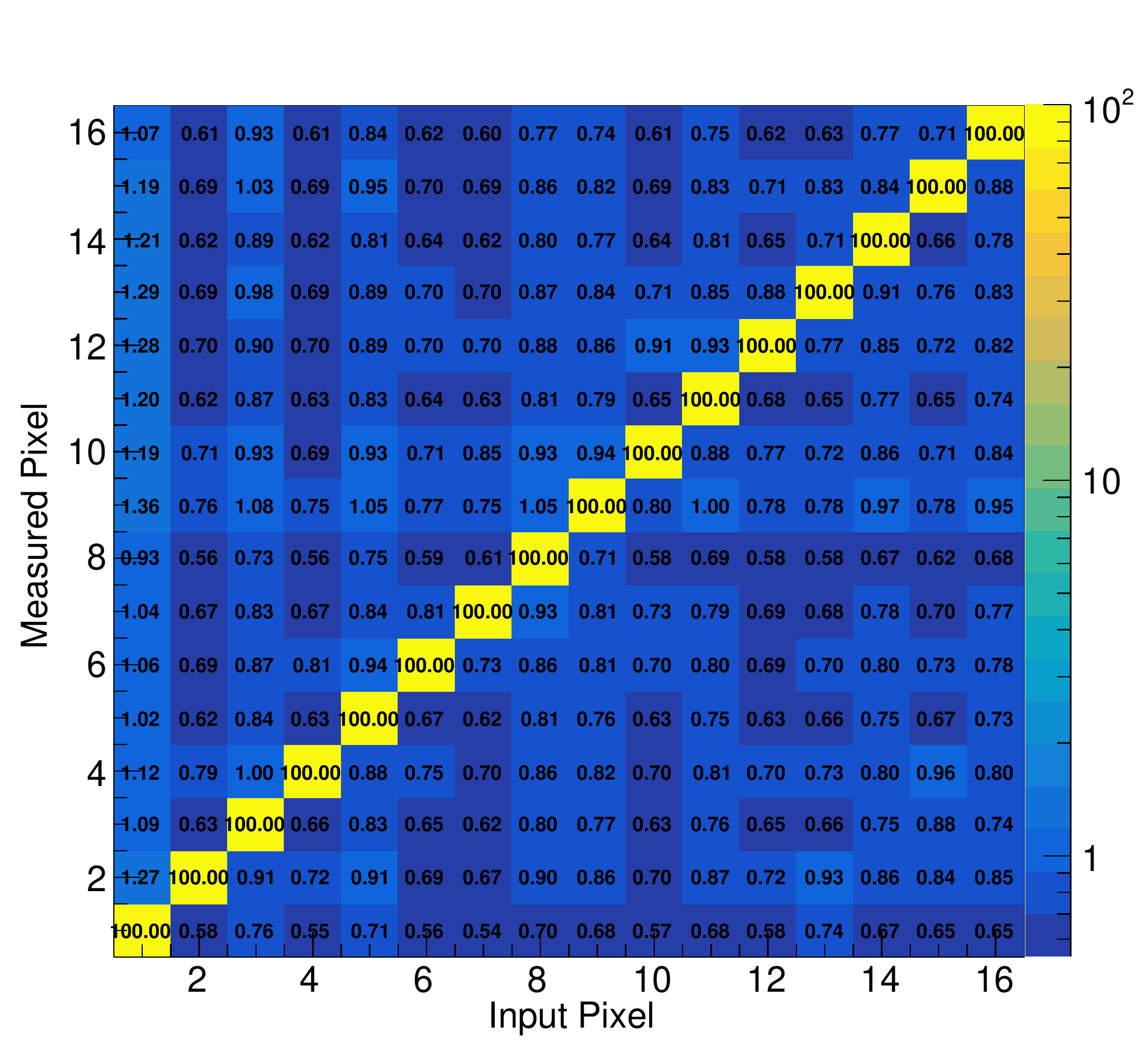}
    \caption{}
  \end{subfigure}
\caption{(a) Crosstalk Integral Calculation.
         (b) Crosstalk Pulse Height Calculation.}
\label{fig:osmo2_th2d}
\end{figure}

The crosstalk percentage for both the pulse height and total integral is defined as the
mean average of each channel's response divided by the mean average of the
target channel's response. A series of 10k events are recorded when the laser
is targeted at a single input pixel. Two example pulse height
distributions for one target pixel position are shown
in~\fig{fig:crosstalkExample}. This procedure is performed using all 16 channels
within the array as the target pixel; the results for both pulse height and
waveform integral are shown in~\fig{fig:osmo2_th2d}

The input pixel in Fig.~\ref{fig:osmo2_th2d} is defined as the target pixel of
the laser. Both integral and pulse height measurements are compared as a
relative percentage of the input pixel. Target channels 2, 4, 6, 7, 10, 12, 13,
and 15 all correspond to one (1$\times$8) column within the SiPM module.

\begin{figure}
\centering
\includegraphics[width=\columnwidth]{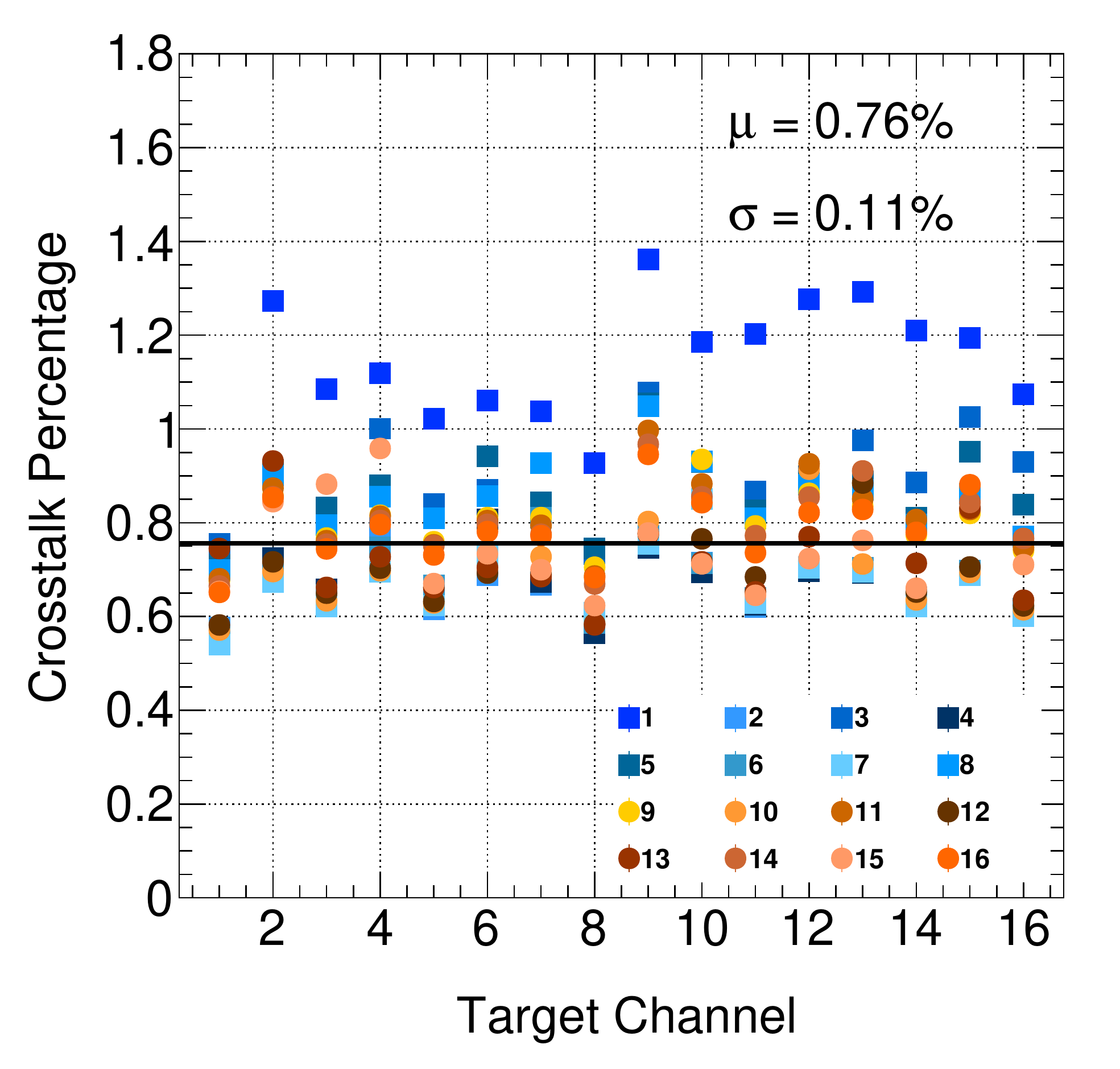}
\caption{Distribution of all of the crosstalk in percentages for pulse 
height. For the zeroth-order fit, we ignore the crosstalk contributions from 
channel 1, since channel 1 had a known electrical calibration error which 
skews the results. Statistical error bars are too small to see on the plot.}
\label{fig:osmo2_distribution}
\end{figure}

We find that the total crosstalk among channels is $\approx$ 1\% or less even
with relatively large input pulses ($\approx$ \SI{200}{mV}) on target
channels.~\fig{fig:osmo2_distribution} shows the pulse height crosstalk
measurement for all channels. The horizontal line  is a 0th-order polynomial fit
to all data, indicating an average crosstalk of 0.76\% for the SiPM module.
Since pulse heights used in detector calibration are in the range
\SIrange{15}{30}{mV}, a \SI{1}{\percent} crosstalk would imply a pulse height (from only
electrical crosstalk) of \SIrange{150}{300}{\micro V}. This value is within 
the measured electrical noise (\SI{\approx0.5}{mV}), and we therefore do not
expect electrical crosstalk to significantly impact our results.

%% OSMO 2 crosstalk distribution

%****************************************************************************************
\subsection{SiPM timing response}

Additional laser studies were performed to characterize the timing resolution as
a function of measured pulse height. In order to vary the laser intensity on the SiPM, an
variable optical attenuator (shown on right image of \fig{fig:laser_side}) was used.
Similar to the electrical crosstalk measurements, laser pulses are sent at a
rate of \SI{50}{Hz} to the SCEMA-B, preceded by a trigger from the DG-535. The time
resolution is then defined as the standard deviation from a Gaussian fit to the time
difference measured between the arrival time of the input trigger and the
rise-time of the pulse height.

\begin{figure}
\centering
\includegraphics[width=\columnwidth]{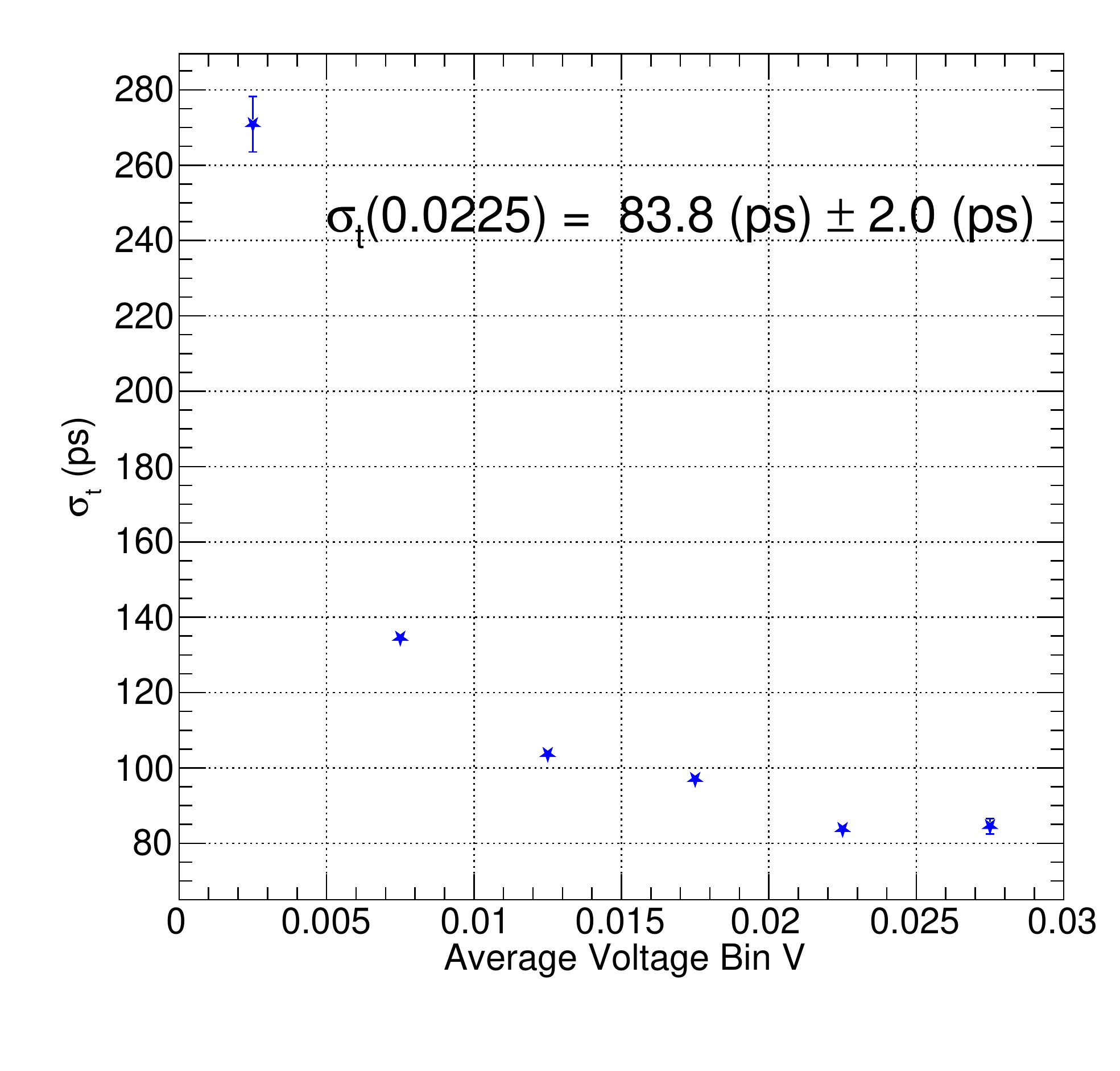}
\caption{Example timing resolution of a single SiPM as a function of measured SiPM pulse
         height. A Photek laser with an optical attenuator is used as an input to a
         SiPM\@. The timing resolution is sigma of a Gaussian fit to the difference in
         time between the trigger and the pulse height. The error bars presented are statistical from the Gaussian fits.}
\label{fig:osmoEnergyTiming}
\end{figure}

Fig.~\ref{fig:osmoEnergyTiming} shows the the approximate timing uncertainty for an example
channel as a function of voltage response from the SiPM due to the laser input\@.
Each point represents time differences with pulse height values measured within \SI{25}{mV} of the
plotted point. We find that, as the pulse height increases, the
timing resolution improves asymptotically to \SI{\approx 80}{ps}. Since the
reported timing jitter between the laser's trigger and pulse is much less
($T_{sl} = \SI{2}{ps}$~\cite{lpg405}), we find that most of the timing
uncertainty is due to the pulse height and readout electronics.

%% include timing description of laser results here

The timing resolution values for small voltages ($<$ \SI{5}{mV}) have large
values and indicate that the SiPM's rise-time compared to noise is not well
measured. Larger values for the SiPM response were not measured since pulses
greater than $\approx$ \SI{30}{mV} are not used in the calibration procedure
described below.

%****************************************************************************************
\subsection{Interaction calibration setup}

The bar calibration procedure involves a single \Natt source. We use the
simultaneous and back-to-back~\SI{0.511}{MeV} gammas using a procedure similar
to methods in both~\cite{galindo-tellez} and~\cite{Sweany}. The tag
scintillator is a teflon-wrapped stilbene crystal that is~\SI{5.0}{cm} in height and has a
\SI{0.5}{cm} $\times$ \SI{0.5}{cm} cross section. The tag scintillator rests on
top of the J-series SensL eval board SiPM and is optically coupled using EJ-550
optical grease.

%% setup figure
\begin{figure}
\centering
\includegraphics[width=\columnwidth]{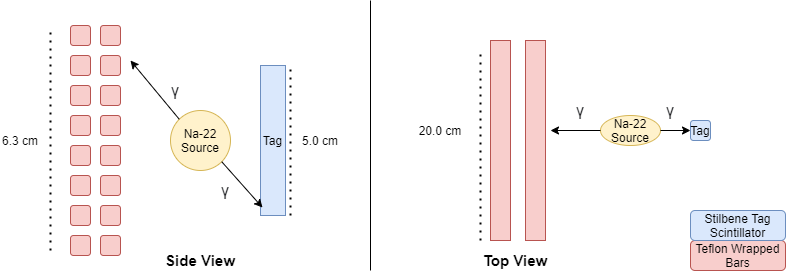}
\caption{Setup of the \Natt tag scintillator and OSMO.}
\label{fig:osmo_setup}
\end{figure}

\begin{figure}
\centering
\includegraphics[width=\columnwidth]{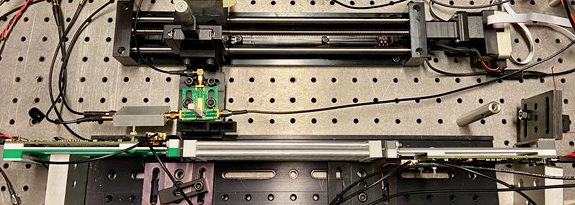}
\caption{Top-down view of the \Natt calibration scan procedure is shown. The
SensL eval board and tag are on the top of the image, with the \Natt source holder
between it and the OSMO.}
\label{fig:osmoExpSetup}
\end{figure}

Since the crystal's height is approximately~\SI{2.0}{cm} smaller than total span
of the eight scintillator bars, the \Natt source is placed closer to the tag to
create a magnification effect so that all of the bars can be targeted by the
range of particles hitting the tag, as shown in~\fig{fig:osmo_setup}. The use of
a larger scintillator tag with magnification allows a single scan across the
length of the OSMO to target the bars in half of the matrix. In order to
precisely move the \Natt source and tag along the length of the OSMO, we use the
eTrack \SI{250}{mm} linear stage  with NEMA 17 MDrive motor from Newmark Systems,
which is controlled via  the host computer for data acquisition through a USB to
RS422 converter. A picture of the experimental setup is shown in~\fig{fig:osmoExpSetup}.

The two independent SCEMA-B boards are powered by a B\&K Precision 9129B DC power
supply, which provides two independent channels of \SI{5.6}{VDC} power, and are
connected to a control computer via micro-USB through a USB hub. The SiPMs are
provided a \SI{30}{V} bias, or an approximate \SI{6}{V} over-voltage, supplied by a B\&K
Precision 1761 power supply. Additionally, \SI{30}{VDC} is supplied to a single
SensL J-series eval board which is used to read out the tag scintillator. We
use both the standard and fast outputs from the eval board. The standard output
is sent to a Photek PA200-10 (Mini-Circuits ZFL-1000LN+) \SI{20}{dB} amplifier, which
is routed to the DG535 pulse generator as the system trigger.

We set a \SI{0.18}{V} input trigger threshold on the DG535. When the threshold is
exceeded, the DG535 issues two parallel triggers (\SI{200}{mV} step function) with
\SI{5}{ps} jitter~\cite{dg535} to the Trigger-A of each SCEMA-B\@. The fast
output from the eval board is sent to the Trigger-B of one SCEMA-B in order
to digitize the fast-output spectrum from the tag and thus measure both an absolute
reference time for the interaction and the tag's \Natt pulse height spectrum.

%****************************************************************************************
\subsection{Interaction energy response}

\begin{figure}
\centering
  \begin{subfigure}{\columnwidth}
    \centering
    \includegraphics[width=\columnwidth]{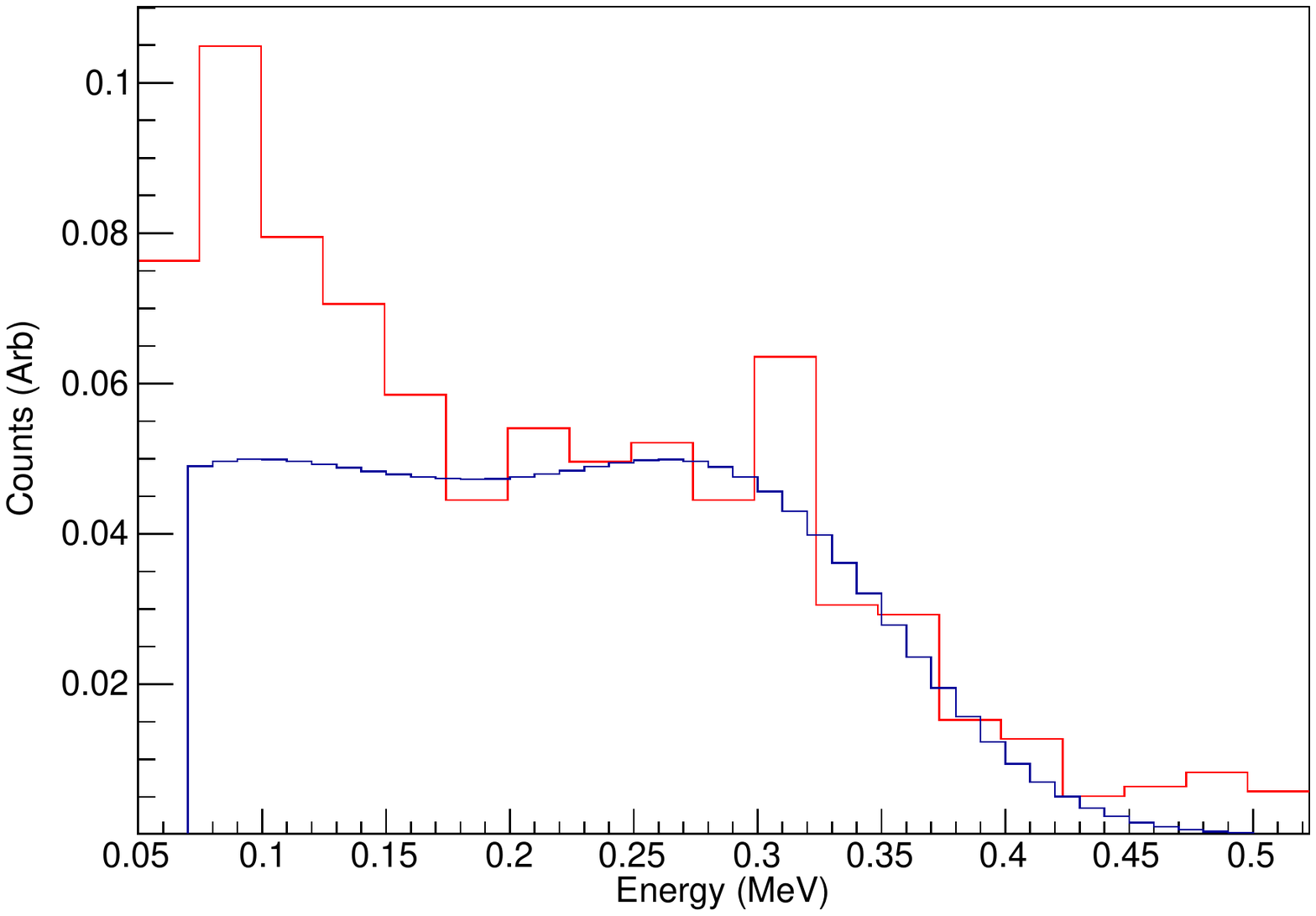}
    \caption{}
  \end{subfigure}
  \begin{subfigure}{\columnwidth}
    \centering
    \includegraphics[width=\columnwidth]{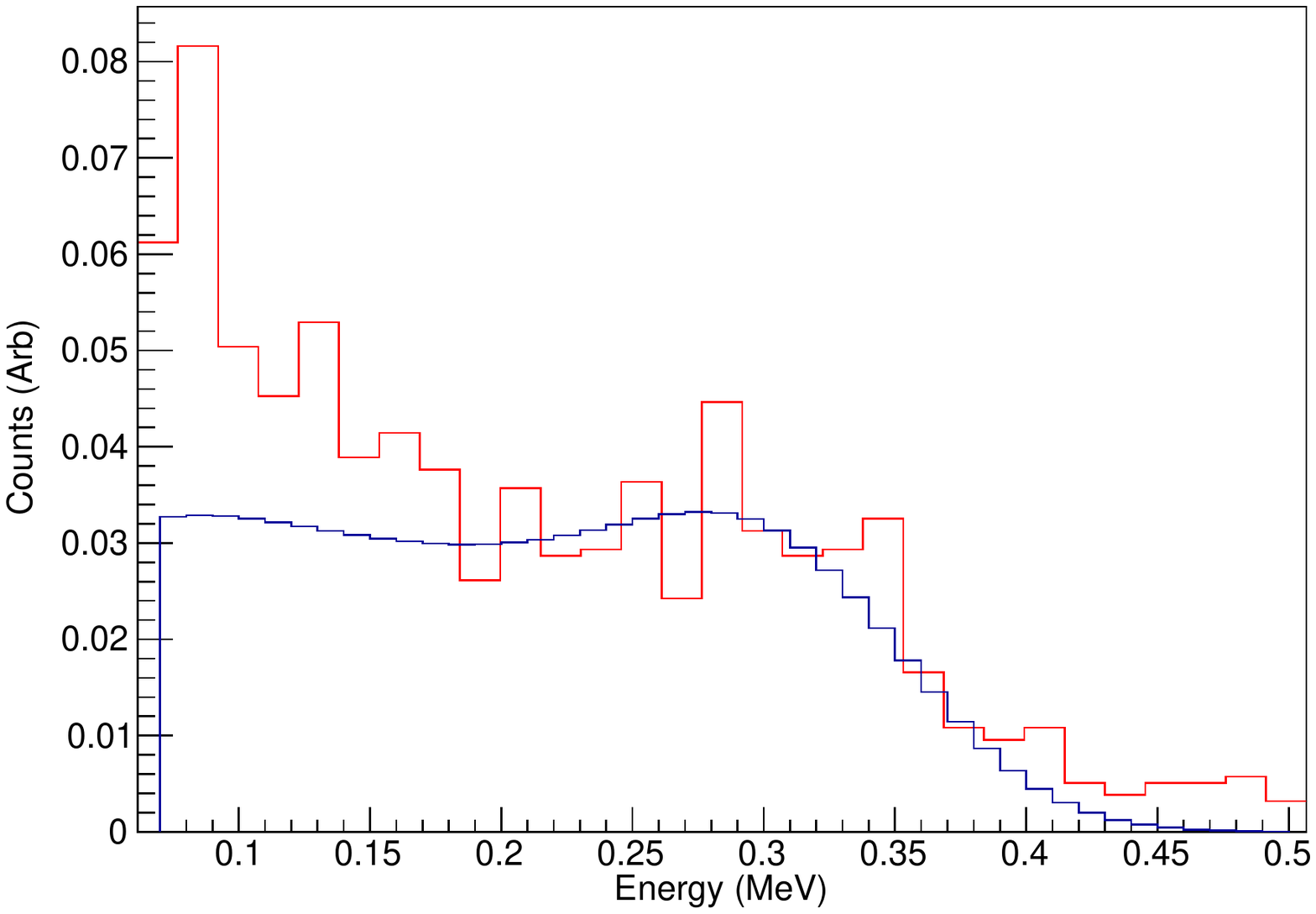}
    \caption{}
  \end{subfigure}
  % \begin{subfigure}{0.32\textwidth}
  %   \centering
  %   \includegraphics[width=\columnwidth]{Figures/osmoBestBar1_energyCalibrationA.pdf}
  %   \caption{}
  % \end{subfigure}
\caption{Example of an energy calibration for a single bar.
The red and blue histograms represent data and the fit respectively.
The region of interest for the Klein-Nishina fit is \SIrange{0.225}{.4}{MeVee}.
(a)/(b) Shows the fit to the Klein-Nishina prediction for the two 
photodetectors on each bar end.}
\label{fig:energyCal}
\end{figure}

The energy spectrum for each bar is fit in the range of \SIrange{0.225}{0.4}{MeV}
to the predicted Klein-Nishina spectrum
for back-to-back \SI{0.511}{MeV} gammas using the method described
in~\cite{Sweany}. This energy calibration is performed for the two
photodetectors coupled to the ends of each bar; an example is shown
in~\fig{fig:energyCal}. The resulting scaling factors for each of the
photodetectors is used to convert measured pulse height in mV to an energy in
MeVee.

%% three canvas of the example energy plots

%% Energy power law fit description here
In the following sections we present the position and time resolution of the bars
for events within different energy ranges and fit the results to a power law.
We quote the position and time resolution energy ranges from
\SIrange{0.0}{0.4}{MeVee} in steps of \SI{0.10}{MeVee}, which fully covers the
Compton spectrum resulting from  \SI{0.511}{MeVee} gammas from the \Natt~source.

%****************************************************************************************
\subsection{Interaction position resolution}

The interaction position along a bar is determined using both the difference
in rising edge time between the two readout SiPMs for each bar and their
relative pulse heights. We combine the measurements using the
Best-Linear-Unbiased-Estimator (BLUE)~\cite{BLUE} to obtain an overall position
resolution.

\begin{figure}
\centering
  \begin{subfigure}{\columnwidth}
    \centering
    \includegraphics[width=0.9\columnwidth,height=0.4\textheight,keepaspectratio]{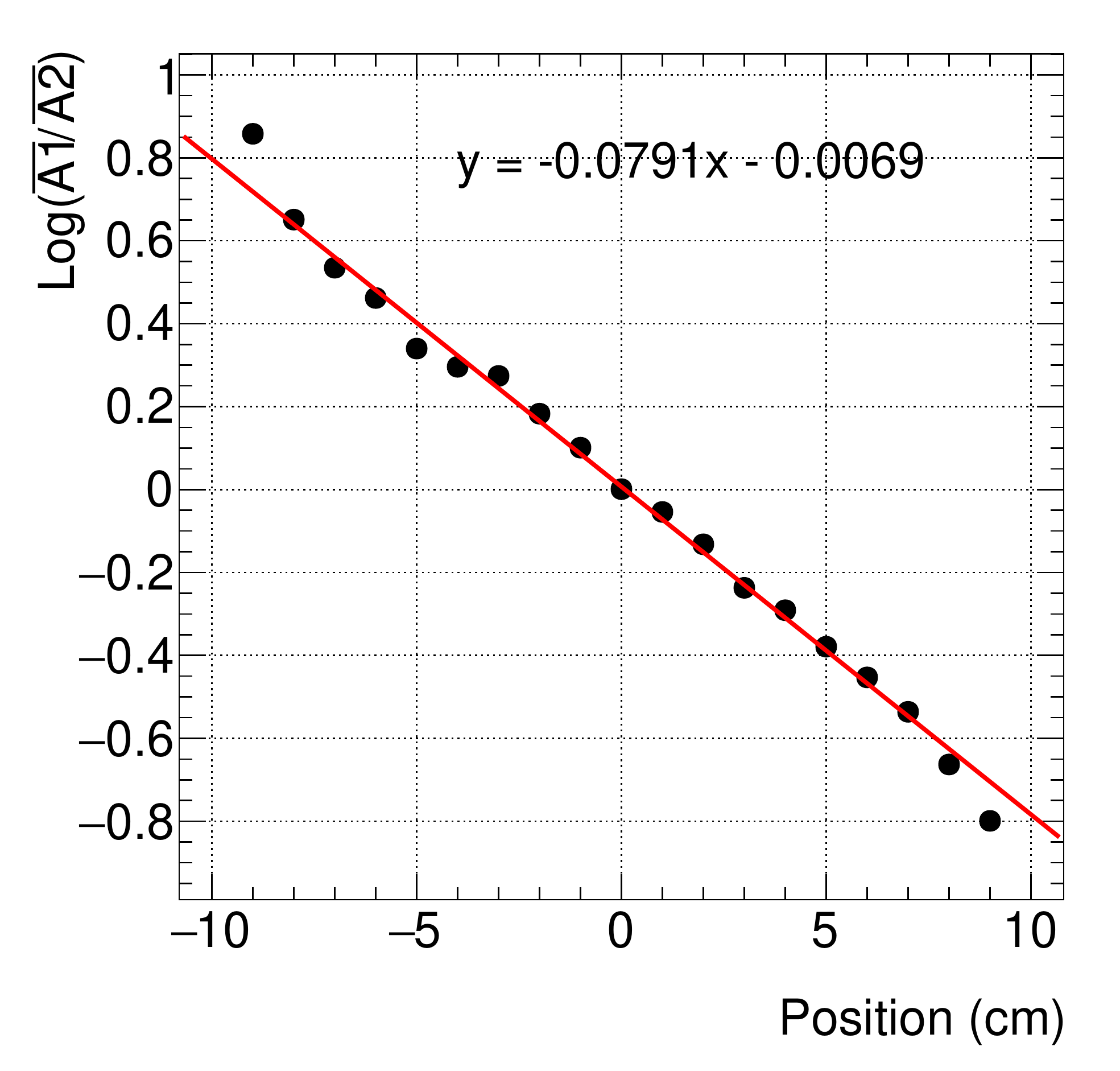}
    \caption{}
  \end{subfigure}
  \begin{subfigure}{\columnwidth}
    \centering
    \includegraphics[width=0.9\columnwidth,height=0.4\textheight,keepaspectratio]{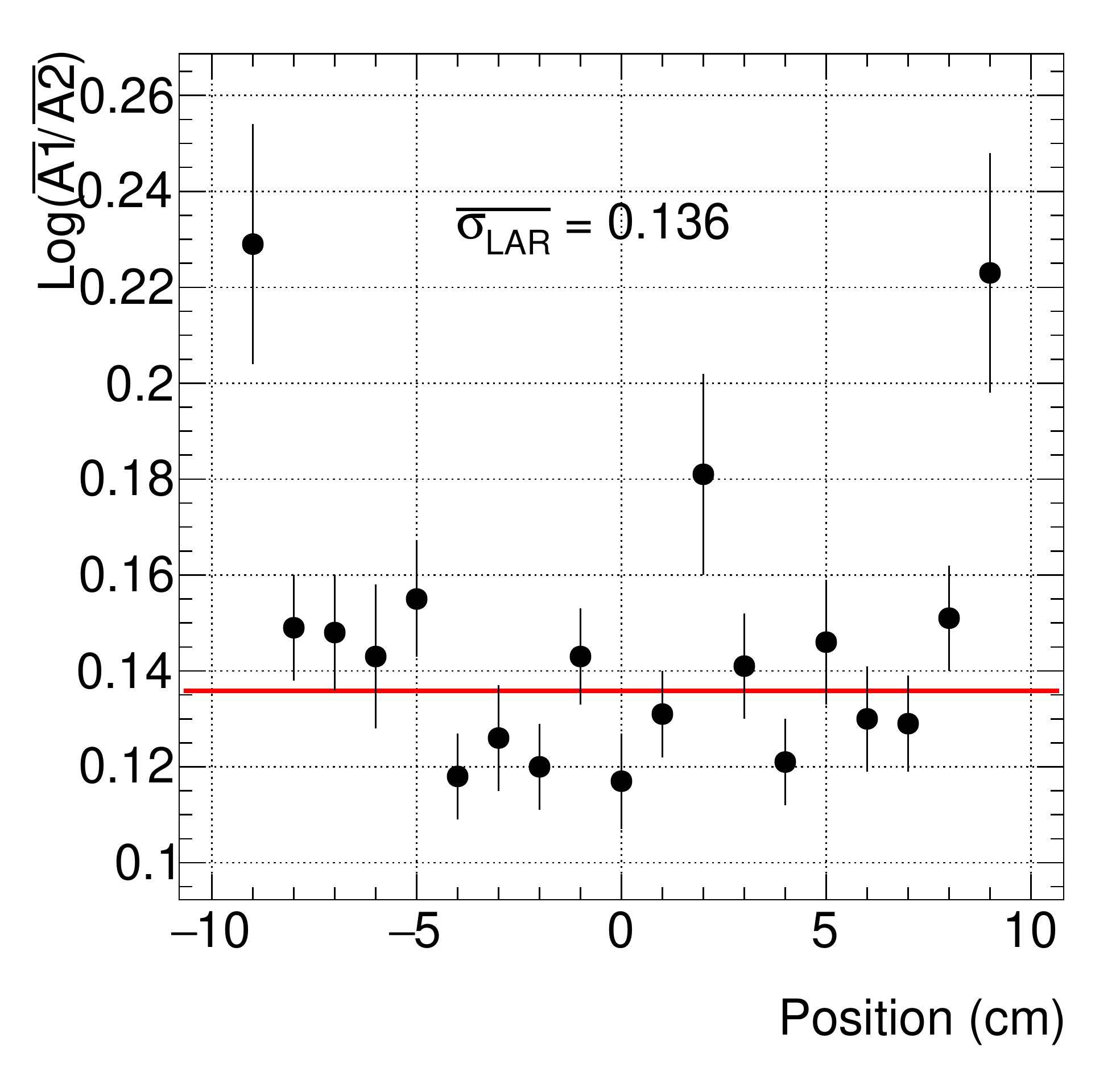}
    \caption{}
  \end{subfigure}
\caption{Example dependence of the mean LAR as a function of source position for Bar 1 in
        the OSMO. After energy calibrations to determine the MeVee/mV ratio for each
        bar, a cut is performed to filter only events in the \SIrange{0.3}{0.4}{MeVee} range.
        Then the LAR is calculated and a Gaussian fit is performed for each source
        position. Data points and uncertainties in (a) are the Gaussian mean and
        uncertainties reported from the (ROOT default) fitting algorithm. In (a), error bars are smaller than the markers. Data points
        and uncertainties in (b) are the fitted Gaussian sigma and the uncertainties
        reported from the fitting algorithm.}
\label{fig:osmo_LAR}
\end{figure}

\begin{figure}
\centering
\includegraphics[width=\columnwidth]{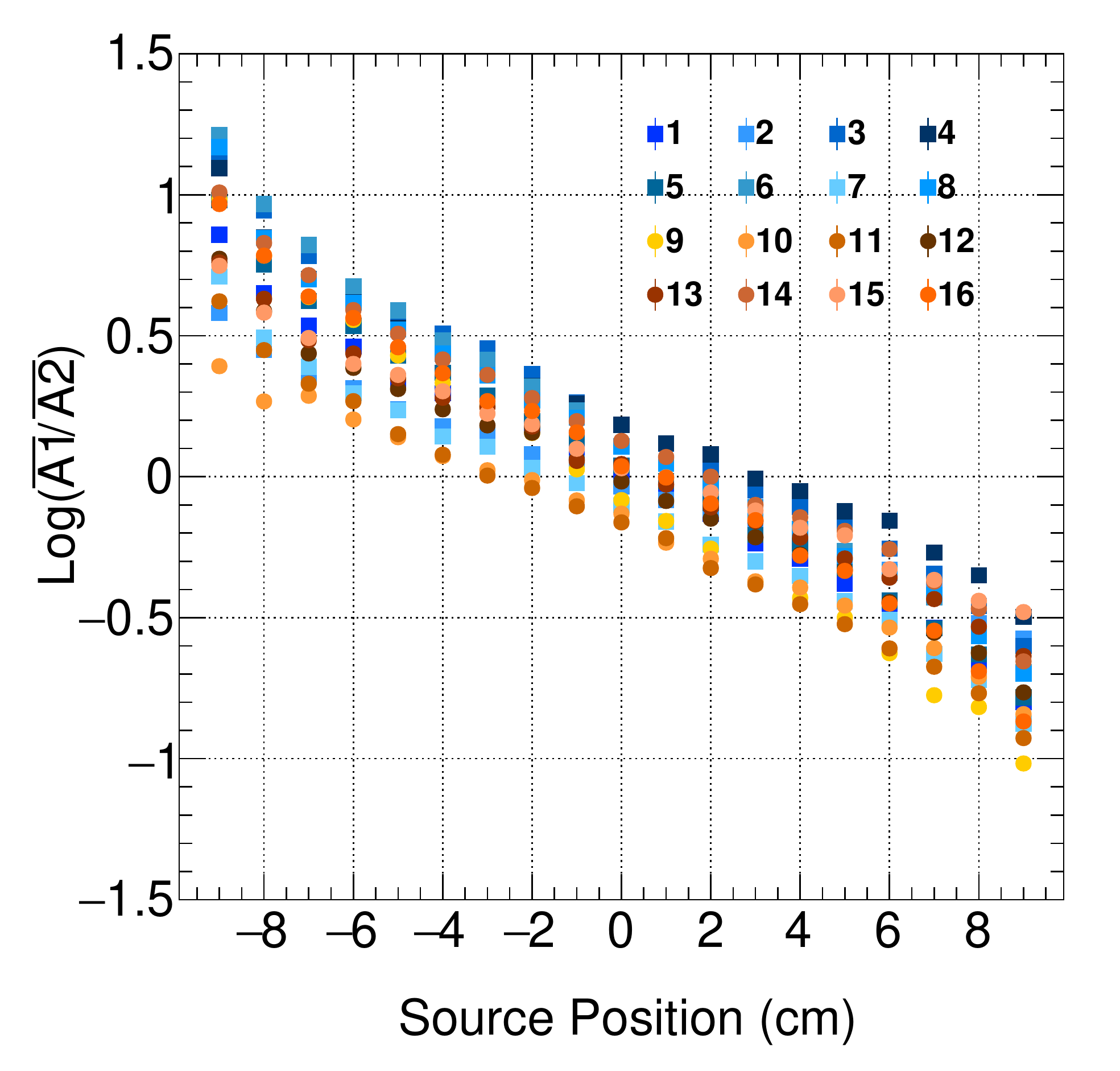}
\caption{The mean LAR for all of the bars as a function of the interaction position.}
\label{fig:osmo_LAR_all}
\end{figure}

The interaction position using SiPM pulse heights from each bar is determined by
the log of the amplitude ratio (LAR) between the two bar ends, or
$\ln \frac{A_{1}}{A_{2}}$, where $A_{1}$ and $A_{2}$ are the pulse heights
measured from the output SiPMs from the end of each bar. The mean LAR as a
function of interaction position for Bar 1 in the \SIrange{0.3}{0.4}{MeVee}
range, measured with a Gaussian fit to the distribution, is shown in
\fig{fig:osmo_LAR}a. The standard deviation of the distribution as a function of
interaction position is shown in \fig{fig:osmo_LAR}b. The reported position
resolution for this bar, which was coupled with optical grease to the SiPM, is
the average LAR width divided by the slope of the LAR means, or 
\SI{1.8}{cm}. \fig{fig:osmo_LAR_all} shows the LAR as a function of interaction
position for all 16 bars coupled with optical grease coupling.

%% Many Bars LAR

%% Many Bars avLAR

In previous studies~\cite{galindo-tellez,Sweany}, the position resolution using
timing was computed by taking the difference of the pulse rise times of each
SiPM from the end of each bar: $t_{1}-t_{2}$. However, in this case the time
measurements on a single OSMO are measured using two different SCEMA-Bs. In these
measurements, the two SCEMA-Bs and their corresponding DRS4s run asynchronously.
We therefore use the time of the synchronized trigger pulse from the
DG-535 into the Trigger-A inputs ($T^A_{1,2}$) in order to align the timing between
SCEMAs:

\begin{equation}
~\label{eqn:delta_time}
\Delta{t} = (T^{A}_{1}-t_{1}) - (T^{A}_{2}-t_{2})
\end{equation}

From the above equation, we substitute $T_{i} = T^{A}_{i} - t_{i}$ to
identify the relative time difference measured on each SCEMA and obtain equation:

\begin{equation}
~\label{eqn:eq_time}
\Delta{t} = T_{1} - T_{2}
\end{equation}

\begin{figure}
\centering
  \begin{subfigure}{\columnwidth}
    \centering
    \includegraphics[width=\columnwidth,height=0.4\textheight,keepaspectratio]{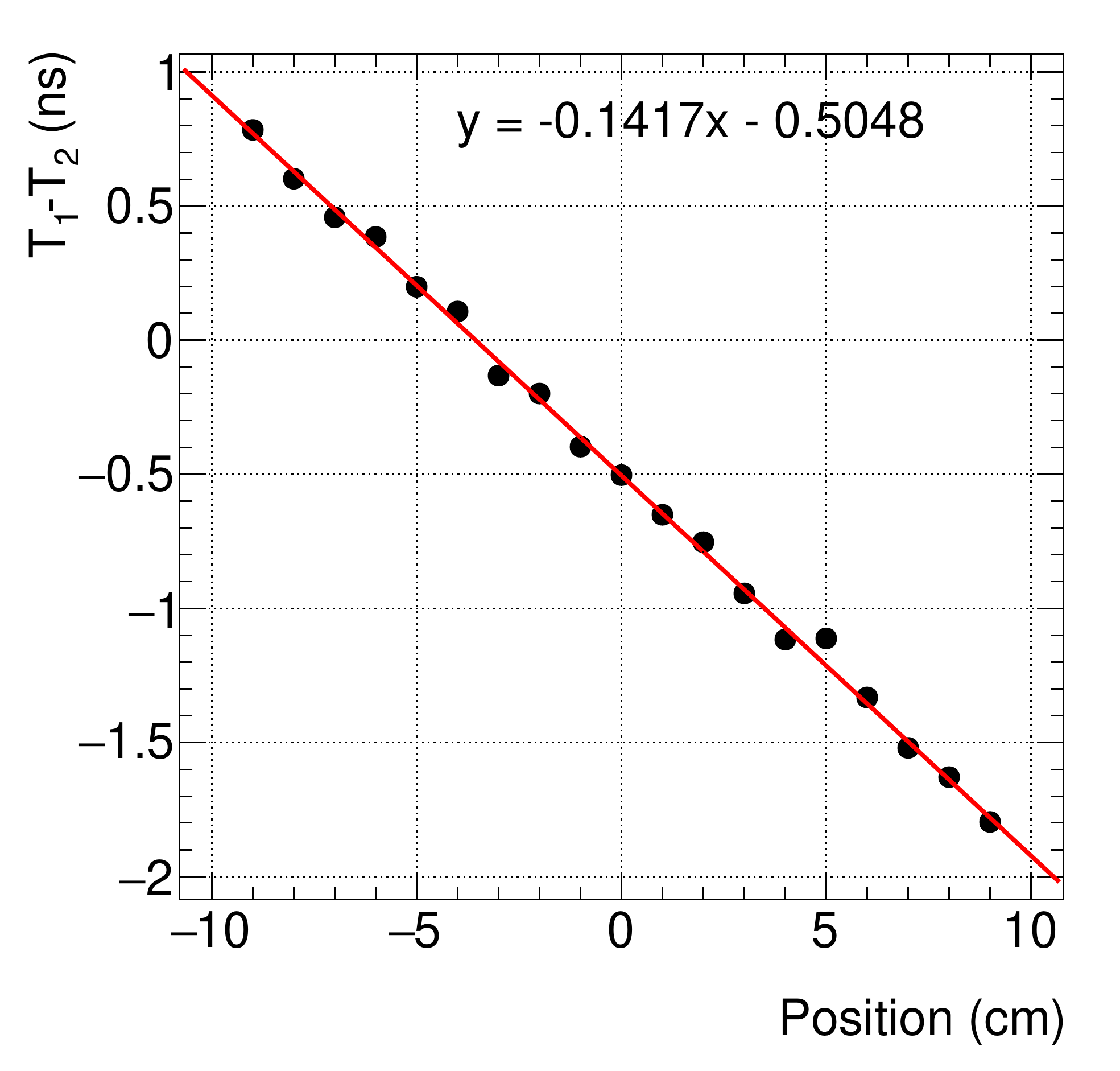}
    \caption{}
  \end{subfigure}
  \begin{subfigure}{\columnwidth}
    \centering
    \includegraphics[width=\columnwidth,height=0.4\textheight,keepaspectratio]{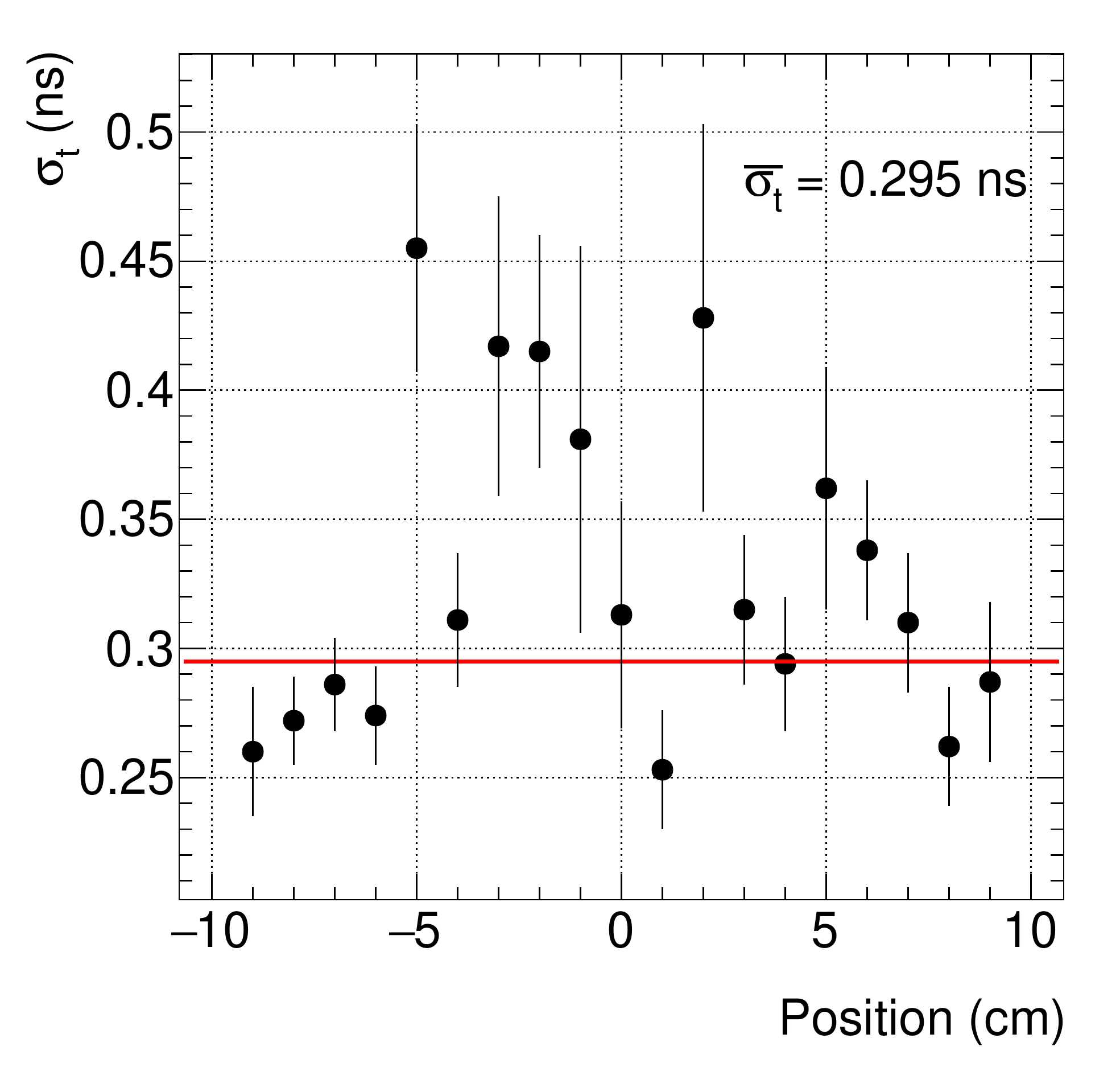}
    \caption{}
  \end{subfigure}
\caption{Example dependence of the time difference as a function of source position for Bar 1 in
        the OSMO. After energy calibrations to determine the MeVee/mV ratio for each
        bar, a cut is performed to filter only events in the \SIrange{0.3}{0.4}{MeVee} range.
        Then the time difference is calculated and a Gaussian fit is performed for each source
        position. Data points and uncertainties in (a) are the Gaussian mean and
        uncertainties reported from the (ROOT default) fitting algorithm. In (a), error bars are smaller than the markers. Data points
        and uncertainties in (b) are the fitted Gaussian sigma and the uncertainties
        reported from the fitting algorithm.
}
\label{fig:osmo_time}
\end{figure}

\begin{figure}
\centering
\includegraphics[width=\columnwidth]{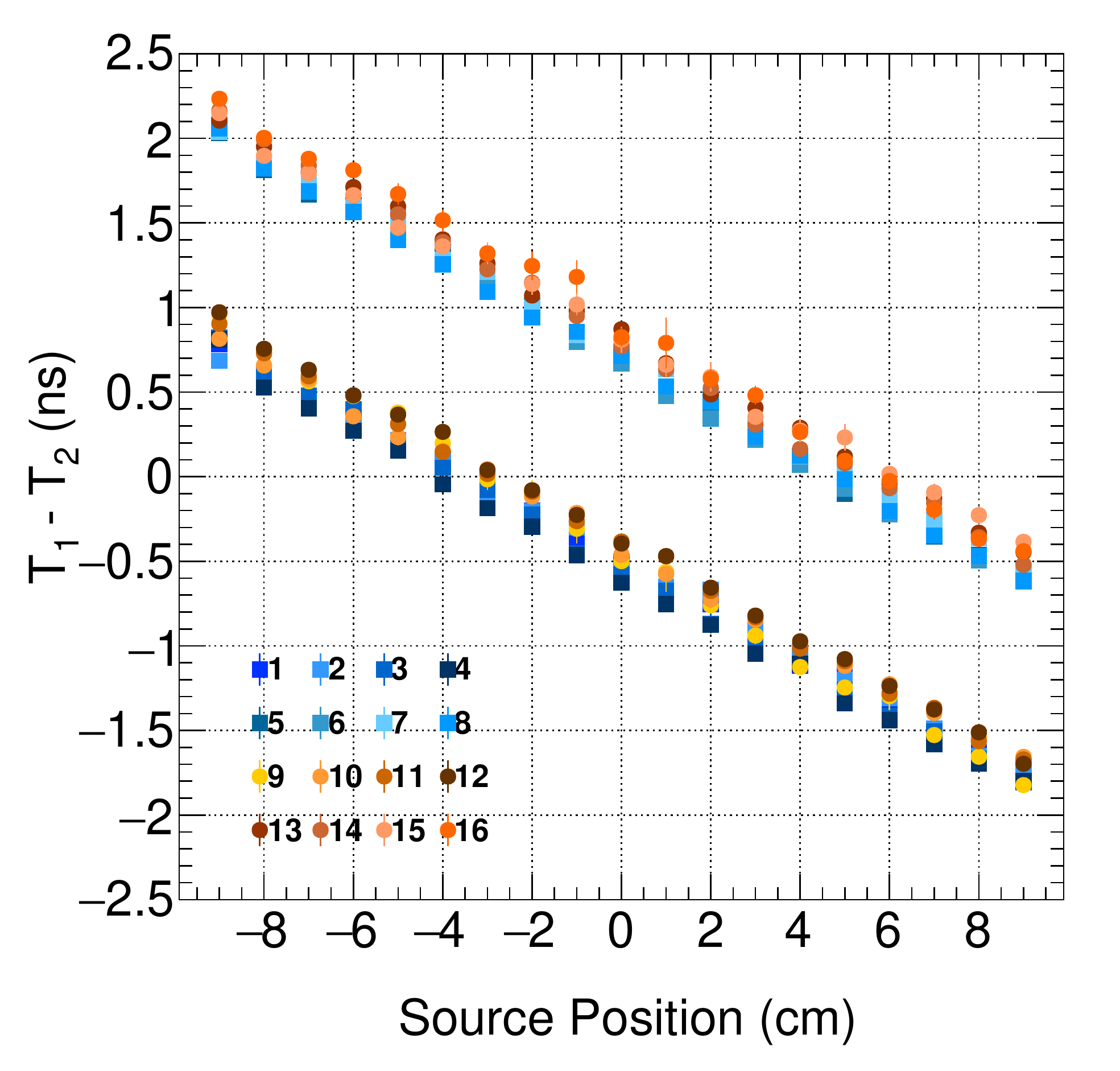}
\caption{The results of a Gaussian fit to the time difference for all bars at 
different \Natt scan positions. The points represent the means and the error 
bars are the sigma from the fit. The two different y-intercept distributions 
correspond to bars digitized by the same DRS4 as the trigger channel (bottom
group) and other DRS4 (top group), and represent the timing offset between the
two DRS4s.}
\label{fig:osmo_time_all}
\end{figure}

\begin{figure}
\centering
\includegraphics[width=\columnwidth]{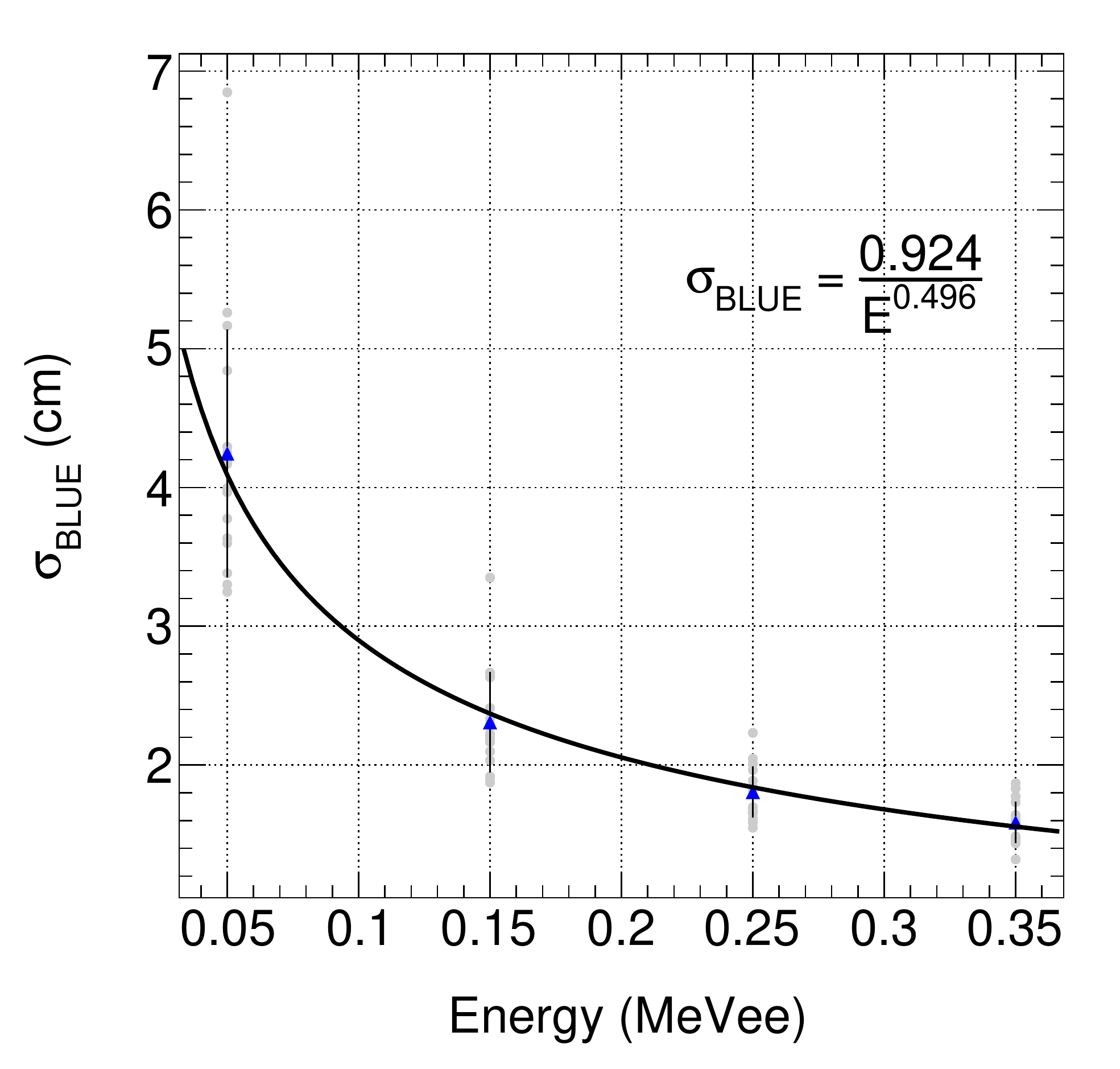}
\caption{The average position resolution for each bar as a function of energy 
is fit to a power law.
Each point's value along the x-axis represents the mean value of the range.
The grey points indicate the results for the 16 bars in the OSMO\@.
The blue points indicate the mean of the 16 bars.
The error bars on the mean data points represent the standard deviation of the 
mean of the 16 bars.
The fit is then applied to the blue points.}
\label{fig:energyResponse}
\end{figure}

\fig{fig:osmo_time} shows the mean and standard deviation of the time difference
distributions as function of interaction position for Bar 1 coupled with optical
grease, again in the \SIrange{0.3}{0.4}{MeVee} range. The position resolution
using time difference for this bar is \SI{2.08}{cm}. \fig{fig:osmo_time_all}
shows the mean of time difference for all 16 bars with optical grease coupling:
the two different y-intercept distributions correspond to the two different DRS4
pairs on each SCEMA-B, resulting in distinct timing offsets. The bottom 8 bars
correspond to when a bar's SiPM channels are read by the same DRS4 as the
trigger channel, and the top 8 bars' SiPM channels are digitized by the other
DRS4. The combined BLUE position resolution as a function of energy is shown in
\fig{fig:energyResponse}, in which the mean position resolution for all 16 bars
is used.

%% example time fits

%% Bar plots for delta-T

%% energy response graph

%****************************************************************************************
\subsection{Interaction time response}

The interaction time is defined as the average of the pulse rise times for
each bar end minus the rise time of the tag pulse ($t_1^B$), which is used as an absolute
time reference. As before, the rise time of each bar end has a correction
from the Trigger-A input to correct for the asynchronous SCEMA acquisition,
including the tag's rise time input to Trigger-B:

\begin{equation}
~\label{eqn:interaction_time}
T_{int} = \frac{(T_{1}^A-t_{1})+(T_{2}^A-t_{2})}{2} - (T_{1}^A-t_{1}^B)
\end{equation}

%% interaction time response single bar example
\begin{figure}
  \begin{subfigure}{\columnwidth}
    \centering
    \includegraphics[width=\columnwidth,height=0.4\textheight,keepaspectratio]{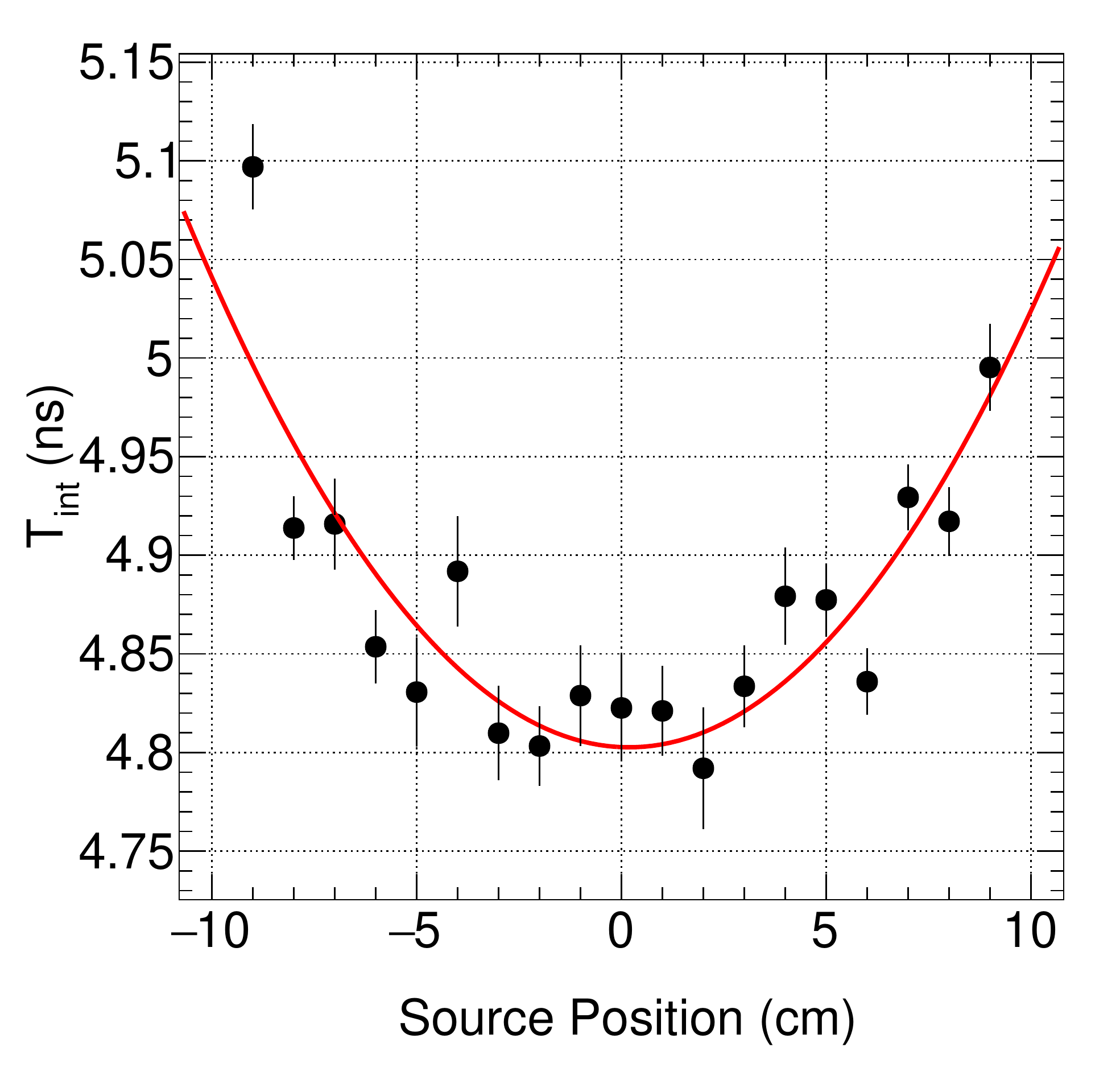}
    \caption{}
  \end{subfigure}
%% average interaction time example for a given bar
  \begin{subfigure}{\columnwidth}
    \centering
    \includegraphics[width=\columnwidth,height=0.4\textheight,keepaspectratio]{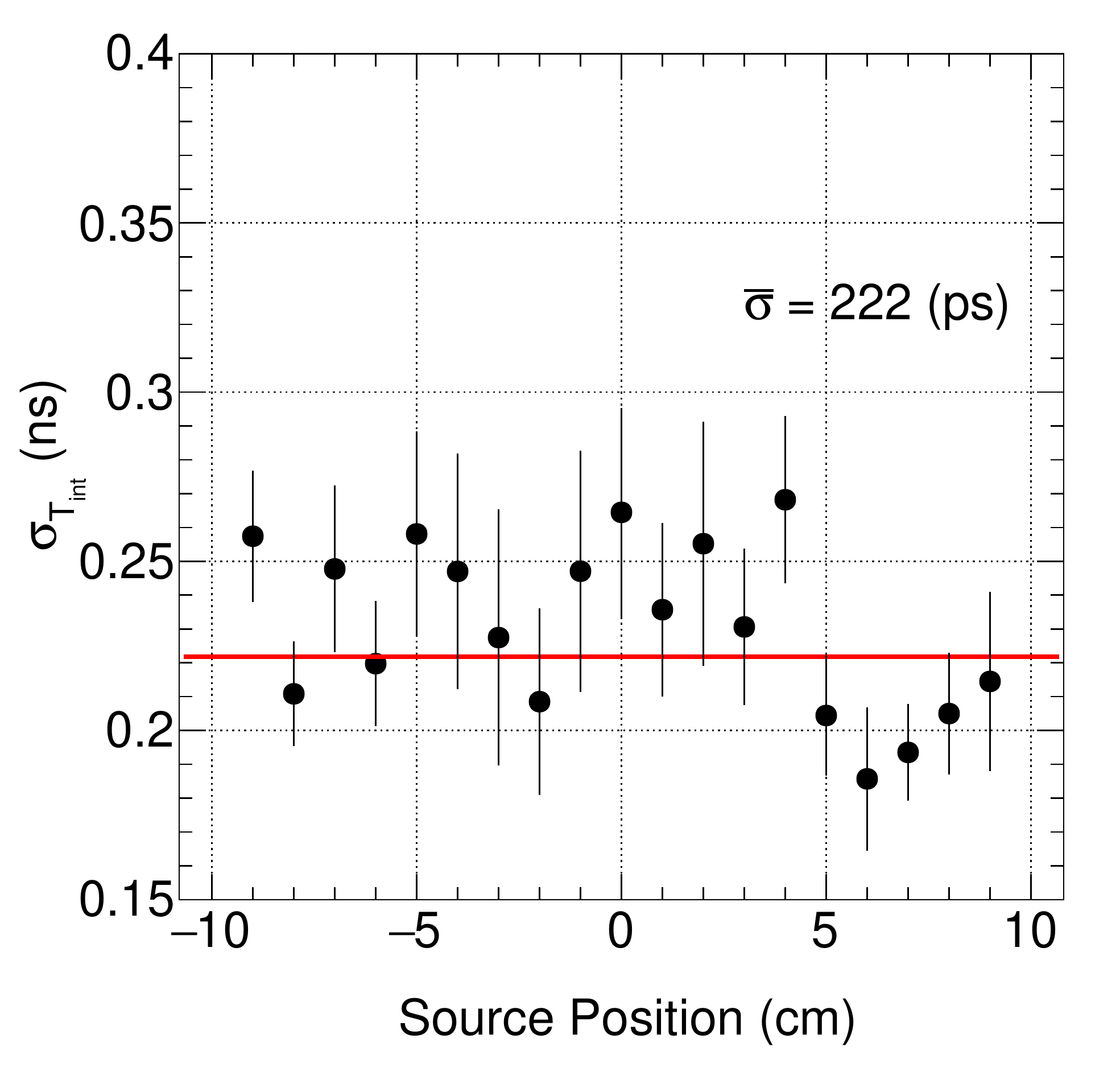}
    \caption{}
  \end{subfigure}
\caption{(a) Interaction time is shown for an example bar (Bar1), in the module.
         We expect a symmetrical effect about the bar length, for any given positional dependence of the interaction time.
         (b) The average uncertainty in the interaction time for Bar 1 at each of the scan positions.
         We fit a zeroth order polynomial to find an average uncertainty for each bar.
        }
\label{fig:osmo_inttime}
\end{figure}

\begin{figure}
\centering
\includegraphics[width=\columnwidth]{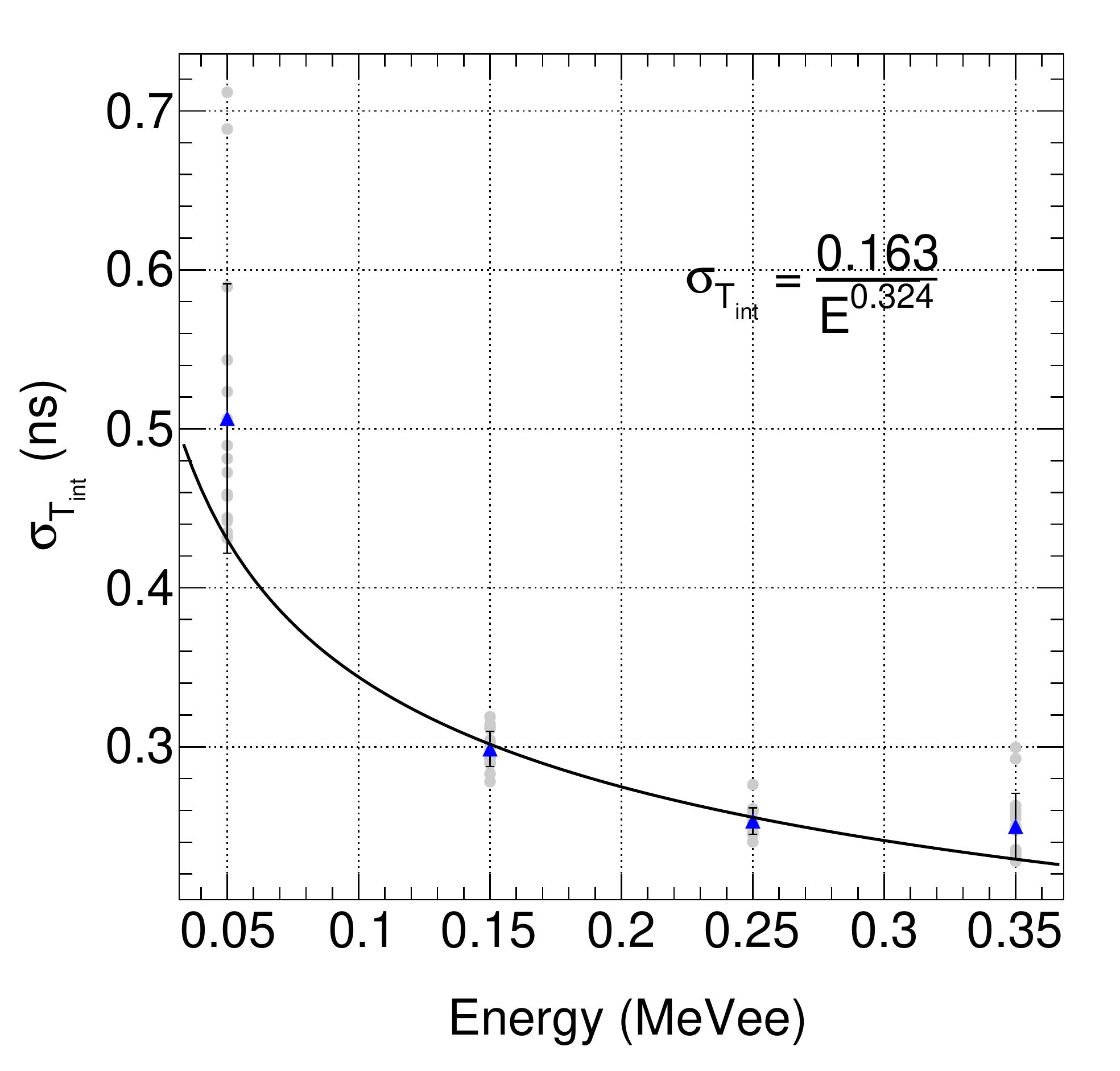}
\caption{The average interaction time for each bar is fit to a power law.
         The grey points indicate the results for the 16 bars in the OSMO\@.
         The blue points indicate the mean of the 16 bars.
         The error bars on the mean data points represent the standard deviation of the
         mean of the 16 bars.
         The fit is then applied to the blue points.
}
\label{fig:interactionTimeResponse}
\end{figure}

The mean and standard deviation of the interaction time distributions as a
function of interaction position for Bar 1 coupled with optical grease are shown
in \fig{fig:osmo_inttime}: the reported interaction resolution is 248 ps for
this bar. It has been previously observed that interaction time within a bar has
a positional dependence along a bar's length~\cite{Sweany},  which is again
observed here. The energy-dependent average interaction time resolution is shown
in \fig{fig:interactionTimeResponse}. The offset from the interaction time
in~\fig{fig:osmo_inttime} occurs from the minimum average delay time from
equation~(\ref{eqn:interaction_time}).

%% interaction time response graph

%****************************************************************************************
\section{Summary of results and discussion}

Here we present position and interaction time resolution results for all 16
channels in the assembled OSMO\@. Tables~\ref{tab:blue_res}
and~\ref{tab:blue_res_may10} summarize the position and timing resolutions for
EJ-550 and EJ-560 optical coupling material, respectively. In the last two rows, we report the mean and
standard  deviation of the given resolution across the 16 bars. We note that
overall, the EJ-550 optical coupling has improved position and timing resolution
values compared to EJ-560. However, there are some bars for which the EJ-560
silicone pads have better position resolution than with optical grease. Compared
to the EJ-550 data, resolution values are noticeably better for bars
\{2,7,15\}, which are close to the tightening screws of the OSMO.

\begin{table}
\caption{Resolution Results with EJ-550 optical grease coupling.}
\label{tab:blue_res}
\setlength{\tabcolsep}{9pt}
\begin{center}
\begin{tabular}{| c | c | c | c | c |}
\hline
Bar ID  &$\sigma_z^{t}$ (cm) &$\sigma_z^{A}$ (cm) & $\sigma_z$ (cm) & $\sigma_{\text{int}}$ (ps)\\
\hline
1 & 2.08 & 1.72 & 1.34 & 222 \\
2 & 2.44 & 2.76 & 1.83 & 244 \\
3 & 1.94 & 1.71 & 1.29 & 242 \\
4 & 1.95 & 2.11 & 1.43 & 233 \\
5 & 1.91 & 1.79 & 1.31 & 226 \\
6 & 1.93 & 1.65 & 1.26 & 227 \\
7 & 2.22 & 2.16 & 1.55 & 228 \\
8 & 2.06 & 1.85 & 1.38 & 231 \\
9 & 1.78 & 1.52 & 1.16 & 234 \\
10 & 2.32 & 2.68 & 1.76 & 247 \\
11 & 2.06 & 2.07 & 1.46 & 229 \\
12 & 1.99 & 2.13 & 1.46 & 239 \\
13 & 2.13 & 2.16 & 1.52 & 224 \\
14 & 2.18 & 1.95 & 1.46 & 238 \\
15 & 2.53 & 2.54 & 1.79 & 263 \\
16 & 1.98 & 1.56 & 1.24 & 240 \\
\hline
$\bar{x}$ & 2.09 & 2.02 & 1.45 & 235 \\
$\sigma_{\text{std}}$ & 0.20 & 0.37 & 0.19 & 10 \\
\hline
\end{tabular}
\end{center}
\end{table}

\begin{table}
\caption{Resolution Results with EJ-560 silicone pads for coupling.}
\label{tab:blue_res_may10}
\setlength{\tabcolsep}{9pt}
\begin{center}
\begin{tabular}{| c | c | c | c | c |}
\hline
Bar ID  &$\sigma_z^{t}$ (cm) &$\sigma_z^{A}$ (cm) & $\sigma_z$ (cm) & $\sigma_{\text{int}}$ (ps) \\
\hline
1 & 2.41 & 2.19 & 1.62 & 253 \\
2 & 2.13 & 2.42 & 1.61 & 244 \\
3 & 3.00 & 4.81 & 2.70 & 258 \\
4 & 2.83 & 4.33 & 2.48 & 264 \\
5 & 2.99 & 4.95 & 2.73 & 262 \\
6 & 2.60 & 3.55 & 2.15 & 297 \\
7 & 2.04 & 1.76 & 1.34 & 238 \\
8 & 2.02 & 1.88 & 1.38 & 248 \\
9 & 2.29 & 2.20 & 1.59 & 263 \\
10 & 7.45 & 9.28 & 5.90 & 352 \\
11 & 2.82 & 3.16 & 2.11 & 264 \\
12 & 3.53 & 6.47 & 3.41 & 300 \\
13 & 3.04 & 2.49 & 1.95 & 280 \\
14 & 2.75 & 2.57 & 1.88 & 250 \\
15 & 2.16 & 1.96 & 1.45 & 237 \\
16 & 2.18 & 2.05 & 1.50 & 236 \\
\hline
$\bar{x}$ & 2.89 & 3.50 & 2.24 & 265 \\
$\sigma_{\text{std}}$ & 1.25 & 1.99 & 1.10 & 29 \\
\hline\end{tabular}
\end{center}
\end{table}

\begin{figure}
\centering
  \begin{subfigure}{\columnwidth}
    \centering
    \includegraphics[width=\columnwidth,height=0.4\textheight,keepaspectratio]{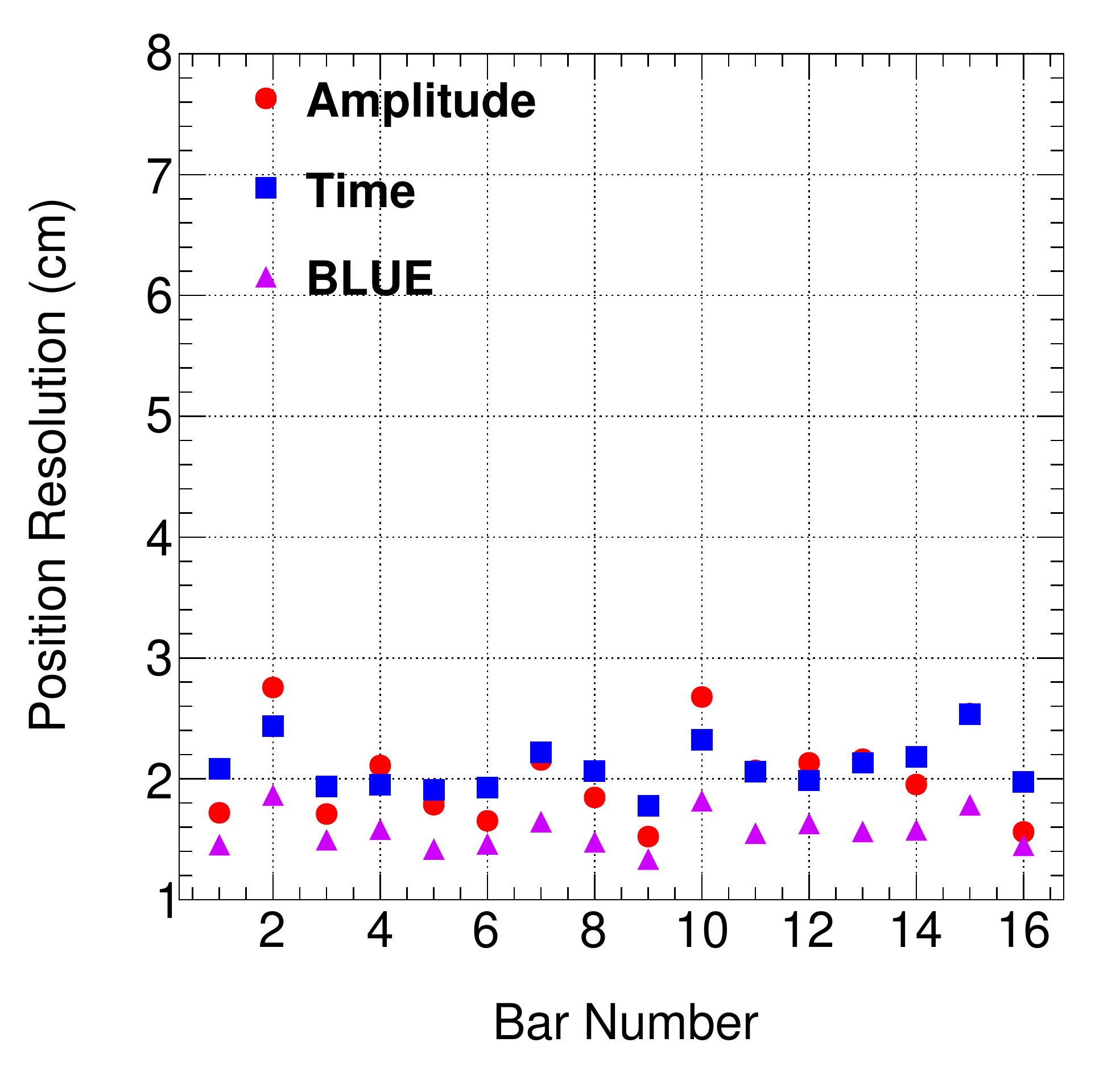}
    \caption{EJ-550 optical grease}
  \end{subfigure}
  \begin{subfigure}{\columnwidth}
    \centering
    \includegraphics[width=\columnwidth,height=0.4\textheight,keepaspectratio]{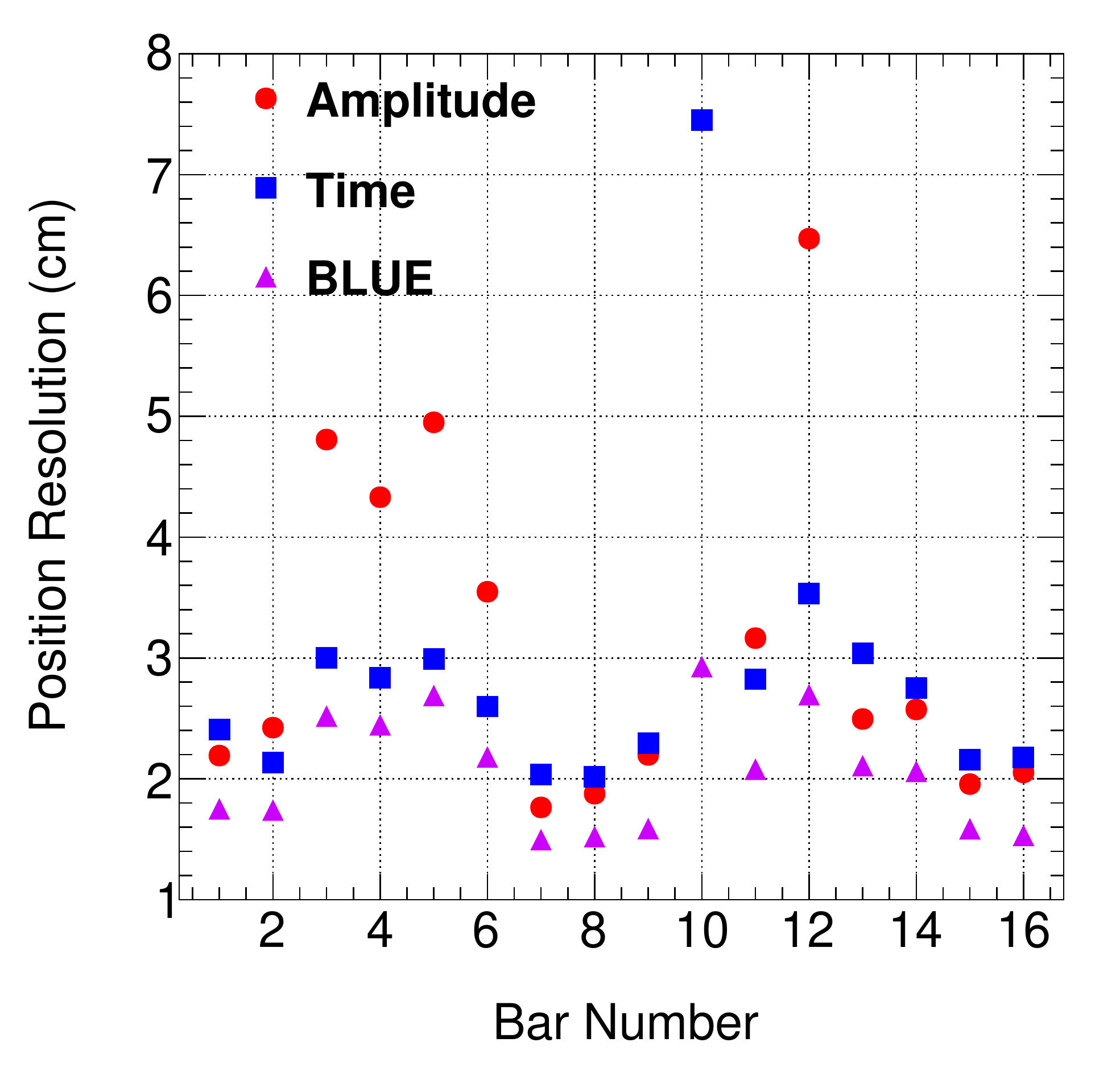}
    \caption{EJ-560 silicone pads}
  \end{subfigure}
\caption{(a) A graphical representation of Table~\ref{tab:blue_res}.
         (b) A graphical representation of Table~\ref{tab:blue_res_may10}.}
\label{fig:resolutions}
\end{figure}

The bars where EJ-560 has improved $\sigma_{\text{blue}}$ resolution compared to EJ-550
are bars 2, 7, and 15. All three of these bars are also close to the tightening screws
on the OSMO\@. One possibility is that these bars have better optical coupling
between the scintillator and SiPM in the optical pad configuration due to better
compression or alignment, and therefore the SiPMs collected more light.
\Fig{fig:resolutions} highlights the
differences in position resolution between the two optical-coupling
configurations.

%% May31 position resolution

%% may31 vs May10 energy scalining
\begin{figure}
\centering
  \begin{subfigure}{\columnwidth}
    \centering
    \includegraphics[width=\columnwidth,height=0.4\textheight,keepaspectratio]{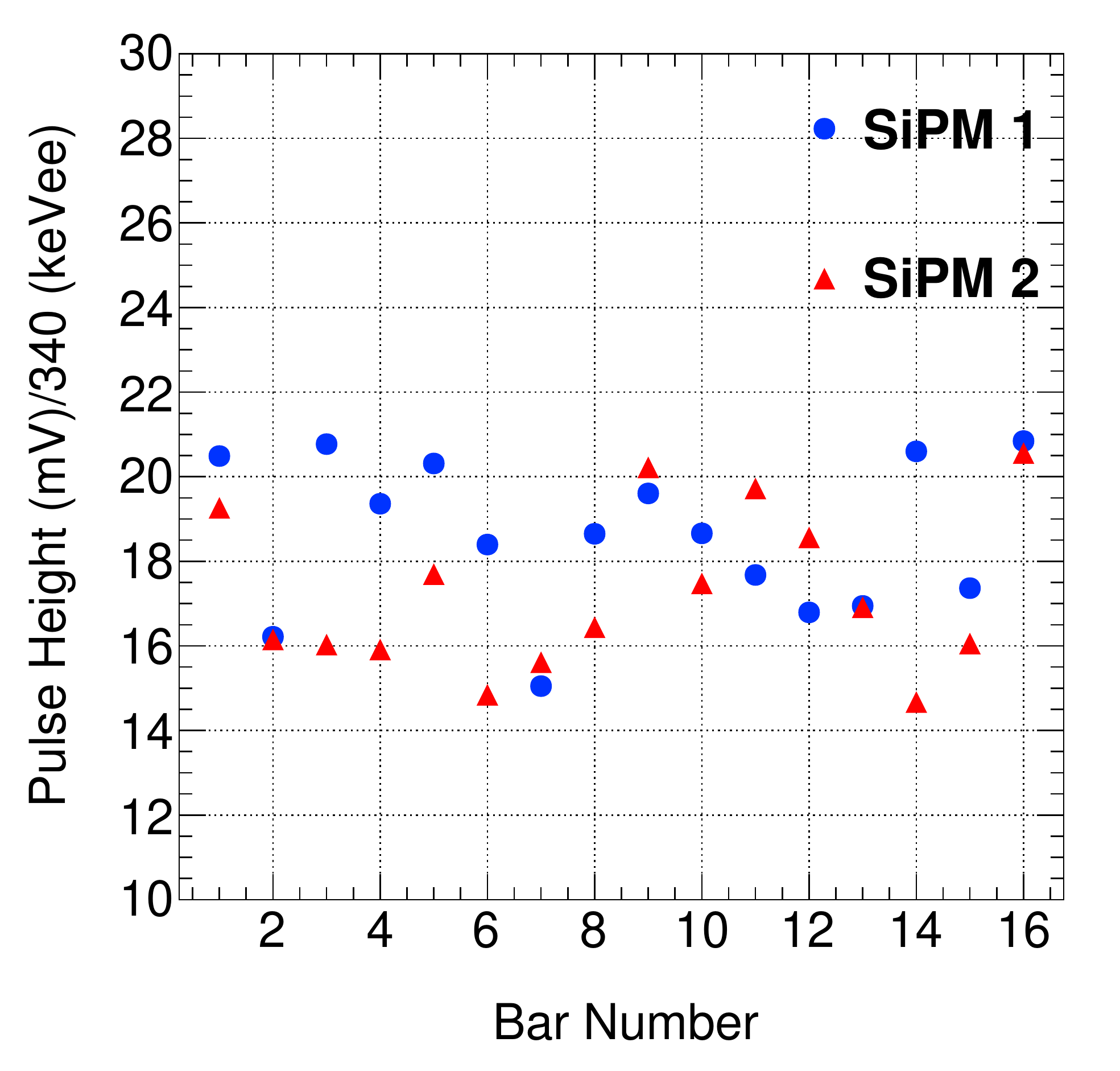}
    \caption{EJ-550 optical grease}
  \end{subfigure}
  \begin{subfigure}{\columnwidth}
    \centering
    \includegraphics[width=\columnwidth,height=0.4\textheight,keepaspectratio]{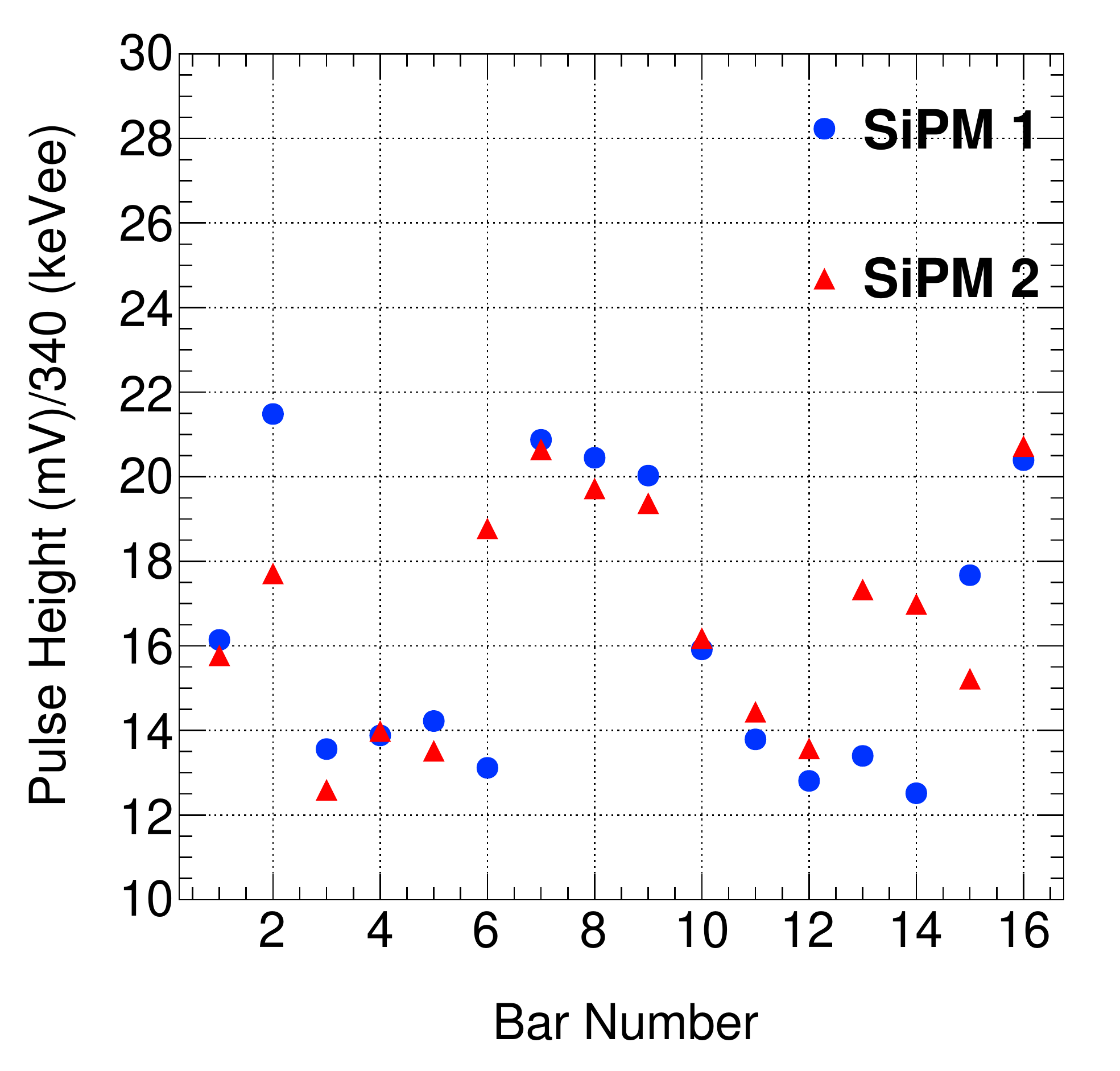}
    \caption{EJ-560 silicone pads}
  \end{subfigure}
\caption{(a) Pulse height response in mV to a \SI{340}{keVee} gamma interaction using the EJ-550 optical grease dataset.
         (b) Pulse height response in mV to a \SI{340}{keVee} gamma interaction using the EJ-560 silicone pad dataset.}
\label{fig:energyScaling}
\end{figure}

We observe the average and standard deviation of position resolution for the 16 bars coupled 
with EJ-550 to be $\SI{1.45}{cm}\pm\SI{0.19}{cm}$ and with EJ-560 to be
$\SI{2.24}{cm} \pm \SI{1.10}{cm}$. We also observe an average and standard
deviation of the interaction time resolution for the 16 bars with EJ-550 to be
$\SI{235}{ps} \pm \SI{10}{ps}$ and with EJ-560 to be
$\SI{265}{ps} \pm \SI{29}{ps}$. We note that bar number 10 had substantially different
position resolution with the EJ-560 coupling, which we also attribute to poor
light collection of the optical pad.

One source of the difference in overall position and interaction time
resolutions between to the two configurations is the amount of light collected
at the ends of the SiPMs. \Fig{fig:energyScaling} highlights the differences in
the results of the energy calibrations for each of the photodetectors for each
bar. The energy calibration results show the amplitude response (in mV)
calculated for the Compton edge of the \Natt source. A decreased amplitude
response corresponds to an overall decrease in the light collected by the SiPM.
A decrease in the amplitude response from a SiPM directly affects both the
timing resolution by affecting SiPM's response time (\fig{fig:osmoEnergyTiming})
and amplitude resolution by reducing the signal-to-noise ratio of the measured
pulse. Therefore, a decrease in the amount of light collected (a lower amplitude
response) will negatively affect both interaction time and position
reconstruction.

Another source of timing uncertainty is the use of multiple DRS4s
which share independent clocks between the two unsynchronized SCEMA-Bs. Previous
studies \cite{Sweany} relied on digitization chains that used channels all on
a single synchronized DRS4. Another potential source of additional timing
uncertainty is the use of \SI{20}{cm} long EJ-204 scintillator bars in this
study compared to \SI{19}{cm} as in previous studies.

Compared to previous results~\cite{Sweany}, we observe an increase in combined
position reconstruction uncertainty from \SI{\approx 1.0}{cm} to 
\SI{\approx 1.45}{cm}.
Additionally, we observe that the interaction time uncertainty has
increased from $\SI{154}{ps}\pm\SI{3}{ps}$ to $\SI{235}{ps}\pm\SI{10}{ps}$.
Position reconstruction and interaction time resolution could be improved both
with increased light detected by a SiPM and in an OSMO where both SCEMAs are
controlled synchronously, which allows all DRS4s to operate with a common clock.
Syncrhonized SCEMA-Bs would eliminate the need to measure $T^{A}_{1,2}$ and remove
the substitution required in~(\ref{eqn:delta_time})
and~(\ref{eqn:interaction_time}).

%****************************************************************************************
\section{Conclusions}

We designed, assembled, and characterized a modular SiPM-based scintillator
array. We also compared detector calibration results for a single module using
both EJ-560 and EJ-550 as optical couplings. We find that, on average, EJ-550
optical grease outperforms the EJ-560 silicone pads, which we attribute to
improved light collection by each SiPM due to improved optical coupling. We
therefore find that a major factor affecting an OSMO's position reconstruction and
interaction time resolutions are the variable optical couplings among the 16
bars. Since the optical coupling affects the amount of light collected by the
SiPM, it also affects the SiPM's timing response which directly affects both the
interaction time uncertainty and the position reconstruction uncertainty.

We suspect improved interaction time resolutions and position reconstruction
would result both from better optical coupling and with synchronized SCEMA-Bs.
Finally, we note that SiPM electronic readout and scintillator differences relative
to~\cite{Sweany} may have also contributed to the differences in both measured
position reconstruction and interaction time resolutions. Plans for future work
include calibrating additional OSMOs, and to combine multiple-synchronized OSMOs
together into a fully realized neutron imaging system. Current neutron-imaging
tests are being performed using the single calibrated module presented here with
the EJ-550 optical grease for coupling.

%****************************************************************************************
%****************************************************************************************
\section*{Acknowledgment}
Sandia National Laboratories is a multimission laboratory managed and operated
by National Technology and Engineering Solutions of Sandia, LLC, a wholly owned
subsidiary of Honeywell International, Inc., for the U.S. Department of Energy's
National Nuclear Security Administration under contract DE-NA0003525. This paper
describes objective technical results and analysis. Any subjective views or
opinions that might be expressed in the paper do not necessarily represent the
views of the U.S. Department of Energy or the United States Government. This
document is approved for relase under release number SAND2021-15772 O.

The authors thank the US DOE National Nuclear Security Administration,
Office of Defense Nuclear Nonproliferation Research and Development for funding
this work.

This material is based upon work supported by the U.S. Department of Energy,
National Nuclear Security Administration through the Nuclear Science and
Security Consortium under Award DE-NA0003180.

%****************************************************************************************
%\bibliographystyle{../../../LatexTools/IEEE_style/IEEEtran}
% \bibliographystyle{IEEEtran}
\bibliographystyle{elsarticle-num}
\bibliography{OSMO}

%\begin{thebibliography}{1}
% No! No more manual bibliography! Run latex, then bibtex, then latex to 
% generate the bib file! Edit OSMO.bib to add new references!
%\end{thebibliography}

\end{document}